\pdfoutput=1

\documentclass[twocolumn,letterpaper,aps,prc,longbibliography,superscriptaddress,showpacs,floatfix]{revtex4-1}

\usepackage{graphicx}	
\usepackage{xspace}     
\usepackage{multirow} 

\newcommand{\pt}{\mbox{$p_T$}\xspace}

\newcommand{\raa}{\mbox{$R_{AA}$}\xspace}

\newcommand{\Npart}{\mbox{$N_{\rm part}$}\xspace}
\newcommand{\Ncoll}{\mbox{$N_{\rm coll}$}\xspace}
\newcommand{\Ncollsq}{\mbox{$N_{\rm coll}^{2}$}\xspace}

\newcommand{\sqrts}{\mbox{$\sqrt{s}$}\xspace}
\newcommand{\sqrtsn}{\mbox{$\sqrt{s_{_{NN}}}$}\xspace}
\newcommand{\jpsi}{\mbox{$J/\psi$}\xspace}

\newcommand{\mumu}{\mbox{$\mu^{+}\mu^{-}$}\xspace}
\newcommand{\pp}{\mbox{$p$$+$$p$}\xspace}
\newcommand{\auau}{\mbox{Au$+$Au}\xspace}
\newcommand{\cucu}{\mbox{Cu$+$Cu}\xspace}
\newcommand{\uu}{\mbox{U$+$U}\xspace}
\newcommand{\cuau}{\mbox{Cu$+$Au}\xspace}

\newcommand{\pbpb}{\mbox{Pb$+$Pb}\xspace}

\newcommand{\yforward}{\mbox{$1.2<y<2.2$}\xspace}
\newcommand{\ybackward}{\mbox{$-2.2<y<-1.2$}\xspace}
\newcommand{\mylessthan}{\mbox{$<$}\xspace}
\begin{document}


\title{Forward $J/\psi$ production in U$+$U collisions 
       at $\sqrt{s_{_{NN}}}$=193~GeV}

\newcommand{\abilene}{Abilene Christian University, Abilene, Texas 79699, USA}
\newcommand{\augie}{Department of Physics, Augustana University, Sioux Falls, South Dakota 57197, USA}
\newcommand{\banaras}{Department of Physics, Banaras Hindu University, Varanasi 221005, India}
\newcommand{\barc}{Bhabha Atomic Research Centre, Bombay 400 085, India}
\newcommand{\baruch}{Baruch College, City University of New York, New York, New York, 10010 USA}
\newcommand{\bnlcoll}{Collider-Accelerator Department, Brookhaven National Laboratory, Upton, New York 11973-5000, USA}
\newcommand{\bnlphys}{Physics Department, Brookhaven National Laboratory, Upton, New York 11973-5000, USA}
\newcommand{\caucr}{University of California-Riverside, Riverside, California 92521, USA}
\newcommand{\charlesczech}{Charles University, Ovocn\'{y} trh 5, Praha 1, 116 36, Prague, Czech Republic}
\newcommand{\chonbuk}{Chonbuk National University, Jeonju, 561-756, Korea}
\newcommand{\ciae}{Science and Technology on Nuclear Data Laboratory, China Institute of Atomic Energy, Beijing 102413, People's Republic of~China}
\newcommand{\cns}{Center for Nuclear Study, Graduate School of Science, University of Tokyo, 7-3-1 Hongo, Bunkyo, Tokyo 113-0033, Japan}
\newcommand{\colorado}{University of Colorado, Boulder, Colorado 80309, USA}
\newcommand{\columbia}{Columbia University, New York, New York 10027 and Nevis Laboratories, Irvington, New York 10533, USA}
\newcommand{\czechtech}{Czech Technical University, Zikova 4, 166 36 Prague 6, Czech Republic}
\newcommand{\elte}{ELTE, E{\"o}tv{\"o}s Lor{\'a}nd University, H-1117 Budapest, P{\'a}zm{\'a}ny P.~s.~1/A, Hungary}
\newcommand{\ewha}{Ewha Womans University, Seoul 120-750, Korea}
\newcommand{\fsu}{Florida State University, Tallahassee, Florida 32306, USA}
\newcommand{\gsu}{Georgia State University, Atlanta, Georgia 30303, USA}
\newcommand{\hanyang}{Hanyang University, Seoul 133-792, Korea}
\newcommand{\hiroshima}{Hiroshima University, Kagamiyama, Higashi-Hiroshima 739-8526, Japan}
\newcommand{\howard}{Department of Physics and Astronomy, Howard University, Washington, DC 20059, USA}
\newcommand{\ihepprot}{IHEP Protvino, State Research Center of Russian Federation, Institute for High Energy Physics, Protvino, 142281, Russia}
\newcommand{\illuiuc}{University of Illinois at Urbana-Champaign, Urbana, Illinois 61801, USA}
\newcommand{\inrras}{Institute for Nuclear Research of the Russian Academy of Sciences, prospekt 60-letiya Oktyabrya 7a, Moscow 117312, Russia}
\newcommand{\instpasczech}{Institute of Physics, Academy of Sciences of the Czech Republic, Na Slovance 2, 182 21 Prague 8, Czech Republic}
\newcommand{\isu}{Iowa State University, Ames, Iowa 50011, USA}
\newcommand{\jaea}{Advanced Science Research Center, Japan Atomic Energy Agency, 2-4 Shirakata Shirane, Tokai-mura, Naka-gun, Ibaraki-ken 319-1195, Japan}
\newcommand{\jyvaskyla}{Helsinki Institute of Physics and University of Jyv{\"a}skyl{\"a}, P.O.Box 35, FI-40014 Jyv{\"a}skyl{\"a}, Finland}
\newcommand{\karoly}{K\'aroly R\'oberts University College, H-3200 Gy\"ngy\"os, M\'atrai \'ut 36, Hungary}
\newcommand{\kek}{KEK, High Energy Accelerator Research Organization, Tsukuba, Ibaraki 305-0801, Japan}
\newcommand{\korea}{Korea University, Seoul, 136-701, Korea}
\newcommand{\kurchatov}{National Research Center ``Kurchatov Institute", Moscow, 123098 Russia}
\newcommand{\kyoto}{Kyoto University, Kyoto 606-8502, Japan}
\newcommand{\labllr}{Laboratoire Leprince-Ringuet, Ecole Polytechnique, CNRS-IN2P3, Route de Saclay, F-91128, Palaiseau, France}
\newcommand{\lahorelums}{Physics Department, Lahore University of Management Sciences, Lahore 54792, Pakistan}
\newcommand{\lawllnl}{Lawrence Livermore National Laboratory, Livermore, California 94550, USA}
\newcommand{\losalamos}{Los Alamos National Laboratory, Los Alamos, New Mexico 87545, USA}
\newcommand{\lund}{Department of Physics, Lund University, Box 118, SE-221 00 Lund, Sweden}
\newcommand{\maryland}{University of Maryland, College Park, Maryland 20742, USA}
\newcommand{\mass}{Department of Physics, University of Massachusetts, Amherst, Massachusetts 01003-9337, USA}
\newcommand{\michigan}{Department of Physics, University of Michigan, Ann Arbor, Michigan 48109-1040, USA}
\newcommand{\muhlenberg}{Muhlenberg College, Allentown, Pennsylvania 18104-5586, USA}
\newcommand{\myongji}{Myongji University, Yongin, Kyonggido 449-728, Korea}
\newcommand{\nagasaki}{Nagasaki Institute of Applied Science, Nagasaki-shi, Nagasaki 851-0193, Japan}
\newcommand{\nara}{Nara Women's University, Kita-uoya Nishi-machi Nara 630-8506, Japan}
\newcommand{\natmephi}{National Research Nuclear University, MEPhI, Moscow Engineering Physics Institute, Moscow, 115409, Russia}
\newcommand{\newmex}{University of New Mexico, Albuquerque, New Mexico 87131, USA}
\newcommand{\nmsu}{New Mexico State University, Las Cruces, New Mexico 88003, USA}
\newcommand{\ohio}{Department of Physics and Astronomy, Ohio University, Athens, Ohio 45701, USA}
\newcommand{\ornl}{Oak Ridge National Laboratory, Oak Ridge, Tennessee 37831, USA}
\newcommand{\orsay}{IPN-Orsay, Univ.~Paris-Sud, CNRS/IN2P3, Universit\'e Paris-Saclay, BP1, F-91406, Orsay, France}
\newcommand{\peking}{Peking University, Beijing 100871, People's Republic of~China}
\newcommand{\pnpi}{PNPI, Petersburg Nuclear Physics Institute, Gatchina, Leningrad region, 188300, Russia}
\newcommand{\riken}{RIKEN Nishina Center for Accelerator-Based Science, Wako, Saitama 351-0198, Japan}
\newcommand{\rikjrbrc}{RIKEN BNL Research Center, Brookhaven National Laboratory, Upton, New York 11973-5000, USA}
\newcommand{\rikkyo}{Physics Department, Rikkyo University, 3-34-1 Nishi-Ikebukuro, Toshima, Tokyo 171-8501, Japan}
\newcommand{\saispbstu}{Saint Petersburg State Polytechnic University, St.~Petersburg, 195251 Russia}
\newcommand{\seoulnat}{Department of Physics and Astronomy, Seoul National University, Seoul 151-742, Korea}
\newcommand{\stonybrkc}{Chemistry Department, Stony Brook University, SUNY, Stony Brook, New York 11794-3400, USA}
\newcommand{\stonycrkp}{Department of Physics and Astronomy, Stony Brook University, SUNY, Stony Brook, New York 11794-3800, USA}
\newcommand{\tenn}{University of Tennessee, Knoxville, Tennessee 37996, USA}
\newcommand{\titech}{Department of Physics, Tokyo Institute of Technology, Oh-okayama, Meguro, Tokyo 152-8551, Japan}
\newcommand{\tsukuba}{Center for Integrated Research in Fundamental Science and Engineering, University of Tsukuba, Tsukuba, Ibaraki 305, Japan}
\newcommand{\vandy}{Vanderbilt University, Nashville, Tennessee 37235, USA}
\newcommand{\weizmann}{Weizmann Institute, Rehovot 76100, Israel}
\newcommand{\wigner}{Institute for Particle and Nuclear Physics, Wigner Research Centre for Physics, Hungarian Academy of Sciences (Wigner RCP, RMKI) H-1525 Budapest 114, POBox 49, Budapest, Hungary}
\newcommand{\yonsei}{Yonsei University, IPAP, Seoul 120-749, Korea}
\newcommand{\zagreb}{University of Zagreb, Faculty of Science, Department of Physics, Bijeni\v{c}ka 32, HR-10002 Zagreb, Croatia}
\affiliation{\abilene}
\affiliation{\augie}
\affiliation{\banaras}
\affiliation{\barc}
\affiliation{\baruch}
\affiliation{\bnlcoll}
\affiliation{\bnlphys}
\affiliation{\caucr}
\affiliation{\charlesczech}
\affiliation{\chonbuk}
\affiliation{\ciae}
\affiliation{\cns}
\affiliation{\colorado}
\affiliation{\columbia}
\affiliation{\czechtech}
\affiliation{\elte}
\affiliation{\ewha}
\affiliation{\fsu}
\affiliation{\gsu}
\affiliation{\hanyang}
\affiliation{\hiroshima}
\affiliation{\howard}
\affiliation{\ihepprot}
\affiliation{\illuiuc}
\affiliation{\inrras}
\affiliation{\instpasczech}
\affiliation{\isu}
\affiliation{\jaea}
\affiliation{\jyvaskyla}
\affiliation{\karoly}
\affiliation{\kek}
\affiliation{\korea}
\affiliation{\kurchatov}
\affiliation{\kyoto}
\affiliation{\labllr}
\affiliation{\lahorelums}
\affiliation{\lawllnl}
\affiliation{\losalamos}
\affiliation{\lund}
\affiliation{\maryland}
\affiliation{\mass}
\affiliation{\michigan}
\affiliation{\muhlenberg}
\affiliation{\myongji}
\affiliation{\nagasaki}
\affiliation{\nara}
\affiliation{\natmephi}
\affiliation{\newmex}
\affiliation{\nmsu}
\affiliation{\ohio}
\affiliation{\ornl}
\affiliation{\orsay}
\affiliation{\peking}
\affiliation{\pnpi}
\affiliation{\riken}
\affiliation{\rikjrbrc}
\affiliation{\rikkyo}
\affiliation{\saispbstu}
\affiliation{\seoulnat}
\affiliation{\stonybrkc}
\affiliation{\stonycrkp}
\affiliation{\tenn}
\affiliation{\titech}
\affiliation{\tsukuba}
\affiliation{\vandy}
\affiliation{\weizmann}
\affiliation{\wigner}
\affiliation{\yonsei}
\affiliation{\zagreb}
\author{A.~Adare} \affiliation{\colorado} 
\author{C.~Aidala} \affiliation{\losalamos} \affiliation{\michigan} 
\author{N.N.~Ajitanand} \affiliation{\stonybrkc} 
\author{Y.~Akiba} \affiliation{\riken} \affiliation{\rikjrbrc} 
\author{R.~Akimoto} \affiliation{\cns} 
\author{J.~Alexander} \affiliation{\stonybrkc} 
\author{M.~Alfred} \affiliation{\howard} 
\author{K.~Aoki} \affiliation{\kek} \affiliation{\riken} 
\author{N.~Apadula} \affiliation{\isu} \affiliation{\stonycrkp} 
\author{H.~Asano} \affiliation{\kyoto} \affiliation{\riken} 
% ...
\collaboration{PHENIX Collaboration} \noaffiliation

\date{\today}


\begin{abstract}


The invariant yields, $dN/dy$, for $J/\psi$ production at forward rapidity
$(1.2<|y|<2.2)$ in U$+$U collisions at $\sqrt{s_{_{NN}}}$=193~GeV have
been measured as a function of collision centrality.  The invariant yields
and nuclear-modification factor $R_{AA}$ are presented and compared with
those from Au$+$Au collisions in the same rapidity range. Additionally,
the direct ratio of the invariant yields from U$+$U and Au$+$Au collisions
within the same centrality class is presented, and used to investigate the
role of $c\bar{c}$ coalescence. Two different parameterizations of the
deformed Woods-Saxon distribution were used in Glauber calculations to
determine the values of the number of nucleon-nucleon collisions in each
centrality class, $N_{\rm coll}$, and these were found to give
significantly different $N_{\rm coll}$ values. Results using
$N_{\rm coll}$ values from both deformed Woods-Saxon distributions are
presented.  The measured ratios show that the $J/\psi$ suppression,
relative to binary collision scaling, is similar in U$+$U and Au$+$Au for
peripheral and midcentral collisions, but that $J/\psi$ show less
suppression for the most central U$+$U collisions. The results are
consistent with a picture in which, for central collisions, increase in
the $J/\psi$ yield due to $c\bar{c}$ coalescence becomes more important
than the decrease in yield due to increased energy density. For midcentral
collisions, the conclusions about the balance between $c\bar{c}$
coalescence and suppression depend on which deformed Woods-Saxon
distribution is used to determine $N_{\rm coll}$.

\end{abstract}


\maketitle

		\section{Introduction}

The study of \jpsi production in high energy heavy ion collisions is 
motivated by the prediction that \jpsi formation would be suppressed by 
color screening effects in the Quark Gluon Plasma 
(QGP)~\cite{Matsui:1986dk}. But relating \jpsi suppression to the energy 
densities of the hot matter formed in heavy ion collisions is complicated 
by the presence of competing effects that also modify \jpsi production. 
Competing effects~\cite{Brambilla:2010cs} can be divided into cold nuclear 
matter (CNM) effects and hot matter effects. CNM effects are those which 
modify the yield or kinematic distributions of \jpsi produced in a nuclear 
target in the absence of a QGP. They include modification of the parton 
densities in a nucleus~\cite{Hirai:2007sx, Eskola:2009uj, Kovarik:2010uv, 
deFlorian:2011fp, Helenius:2012wd}, breakup of the \jpsi or its $c\bar{c}$ 
precursor state in the nuclear target due to collisions with 
nucleons~\cite{Lourenco:2008sk, Arnaldi:2010ky, McGlinchey:2012bp}, 
transverse momentum broadening due to the $c\bar{c}$ traversing the cold 
nucleus, and initial state parton energy loss~\cite{Arleo:2012hn}.  CNM 
effects are expected to be strongly dependent on collision system, 
collision energy, rapidity and collision centrality. In hot matter, \jpsi 
yields can be enhanced by coalescence of $c\bar{c}$ pairs that are 
initially unbound, but which become bound due to interactions with the 
medium~\cite{Zhao:2010nk}. At sufficiently high charm production rates, 
there can be a significant yield of \jpsi from coalescence of a $c$ and 
$\bar{c}$ from different hard processes.

Precise \jpsi data extending down to zero $p_T$ have been published for 
Pb$+$Pb or Au$+$Au collisions at energies of $\sqrtsn = 17.3$ 
GeV~\cite{Alessandro:2004ap}, 62.4~GeV~\cite{Adare:2012wf,Zha:2014nia}, 
200~GeV~\cite{Adare:2011yf,Adare:2006ns,Adamczyk:2013tvk} and 2.76 TeV 
~\cite{Abelev:2013ila}. The nuclear-modification factor (\raa) of the 
\jpsi yield in the highest centrality collisions is observed to drop from 
$\sqrtsn = 17.3$~GeV to 200~GeV, and then increase strongly between 200 
GeV and 2.76 TeV~\cite{Adare:2011yf,Adare:2006ns,Abelev:2013ila}.  The 
rise in \raa between 200~GeV and 2.76~TeV is well 
described~\cite{Adam:2015isa} in magnitude and transverse momentum 
dependence by models that include coalescence of $c$ and $\bar{c}$ pairs 
from different hard scattering processes~\cite{Zhao:2010nk, Zhou:2014kka}. 
The energy dependence of the modification suggests that \jpsi production, 
after accounting for the modification due to cold nuclear matter 
effects~\cite{Arnaldi:2010ky, Lourenco:2008sk, Brambilla:2010cs}, is 
increasingly suppressed from 17.3~GeV to 200~GeV by stronger color 
screening in the increasingly hot QGP. But when the collision energy 
increases to 2.76~TeV, the rising underlying charm production rate leads 
to an increasing, and eventually dominant, contribution to the \jpsi cross 
section from coalescence.

Of the collision energies observed so far the nuclear modification, \raa, 
for the most central collisions is at a minimum at \sqrtsn = 200~GeV, 
although there is little difference from 62.4~GeV, where the energy 
density is smaller. The behavior of \raa in the range of energy densities 
accessed at energies between \sqrtsn = 200~GeV in \auau collisions and 
2.76~TeV in \pbpb collisions is not known. The measurement of \jpsi yields 
in \sqrtsn = 193~GeV $^{238}{\rm U}{+}^{238}{\rm U}$ 
collisions, the largest system yet studied at 
RHIC, during the 2012 data taking at the Relativistic Heavy Ion Collider 
(RHIC) run provides an opportunity to increase the energy density above 
that for \sqrtsn = 200~GeV \auau collisions by 
$\sim$20\%~\cite{Kikola:2011zz}, and to observe the effect on the measured 
\raa.  In this paper we report the results of measurements of the \jpsi 
yield in \uu collisions at forward and backward rapidity at 
\sqrtsn = 193~GeV using the PHENIX detector.

		\section{PHENIX Detector}

\begin{figure}[!htb]
	\includegraphics[width=1.0\linewidth]{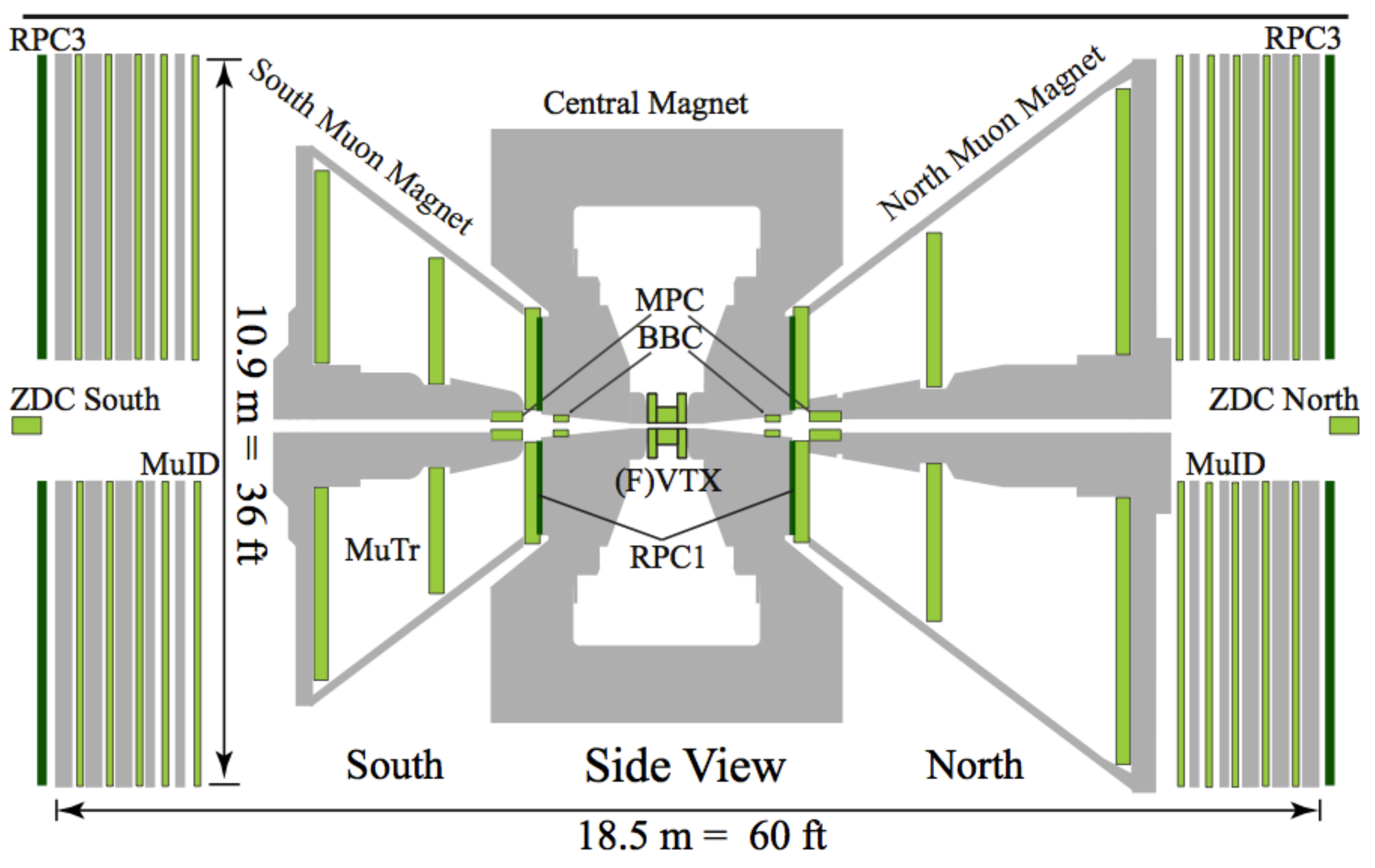}
       \caption{\label{fig:Detector} (Color online)
A schematic side view of the PHENIX muon arms in 2012.  The 
central arms, which are at midrapidity, are not shown.}
\end{figure}

The \uu data used in this analysis were recorded at RHIC during 2012. 
The PHENIX detector~\cite{PHENIX:NIM} configuration used is shown in 
Fig.~\ref{fig:Detector}. The \jpsi yields reported in this paper were all 
recorded in the muon spectrometers~\cite{PHENIX:MuonNIM}.

The minimum-bias trigger was used for this data set. This required two or 
more hits in each arm of the beam beam counter (BBC), which comprises 
two quartz arrays of 64 \v{C}erenkov counters, positioned symmetrically in the pseudorapidity range  
3.0\mylessthan$|\eta|$\mylessthan3.9.  The time difference between the 
two BBC arms provides the $z$-vertex position both in the minimum-bias 
trigger and in the offline analysis.

The \jpsi were reconstructed from their \jpsi$\rightarrow\mu^+\mu^-$ 
decays, with the muons being detected in two muon spectrometer arms that 
cover the rapidity ranges $-2.2<y<-1.2$ (south) and $1.2<y<2.4$ (north). 
Each comprises a magnet, a copper and steel absorber followed by the muon 
tracker (MuTr) and the muon identifier (MuID). They are described in detail 
in Ref.~\cite{Aronson:2003ab}. An additional steel absorber material of 
thickness 36.2\,cm was added in 2010 to improve the muon yield relative to 
the hadronic background. The muon track momentum is measured in the MuTr, 
which has three stations, each comprising two or three chambers with two cathode 
strip planes each, with half of these planes having no stereo angle, and half having 
their cathode planes tilted at stereo angles that vary between 3 and 11.25 degrees.
Discrimination between muons and punch-through hadrons is 
provided by the MuID, which comprises alternating steel absorber layers 
interleaved with Iarocci tubes. A muon candidate is required to penetrate 
all the way to the last layer of the MuID, requiring a minimum muon 
momentum of 3~GeV/$c$. The acceptance for the \jpsi, discussed in detail 
in section~\ref{efficiency}, is flat to within $\approx30\%$ from transverse momentum, \pt, of
zero to 8~GeV/$c$.

		\section{Data Analysis}

For this analysis $1.08 \times 10^{9}$ events with primary vertices within $\pm$\,30\,cm of 
the nominal interaction point were analyzed. This corresponds to an 
integrated luminosity of $\mathcal{L}$ $=$155.6 $\mu$b$^{-1}$ and a 
nucleon-nucleon integrated luminosity of $\int{\mathcal{L_{NN}}dt}$ 
$=238\times238\times155.6$ $\mu$b$^{-1}$ = 8.8 pb$^{-1}$.

Because the invariant yields in the two muon arms must be identical for 
the symmetric \uu collision system, the analysis in the north muon arm was 
restricted to $1.2 < y < 2.2$ to make the rapidity interval equal for the 
north and south arms.

		\subsection{Centrality Determination}
		\label{sec:centrality}

The centrality determination for \uu collisions is based on the combined 
charge sum signals from the two BBC detectors.

The \uu collisions were modeled using a Monte-Carlo simulation based on the 
Glauber model~\cite{Miller:2007ri} with a deformed Woods-Saxon distribution 
for the U nucleus which accounts for its prolate nature.

\begin{equation}
\label{eqn:UUshape1}
\ensuremath{
\rho = \frac{\rho_0}{1 + e^{([r - R^{\prime}]/a)}}
}
\end{equation} 

\noindent
where $R^{\prime}$ depends on the polar angle $\theta$:

\begin{equation}
\label{eqn:UUshape2}
\ensuremath{
R^{\prime} = R [1 + \beta_2 Y^0_2 (\theta) + \beta_4 Y^0_4 (\theta)]
}
\end{equation} 

\noindent
and where $Y^0$ is a Legendre polynomial, $\rho_0$ is the normal nuclear 
density, $R$ is the radius of the nucleus, and $a$ is the surface 
diffuseness parameter.

We considered two parameterizations of the deformed Woods-Saxon 
distribution for U. The first parameter set, which we call set 1, is 
from Masui $et~al.$~\cite{Masui:2009qk}. The second, which we call set 2, 
is from Shou $et~al.$~\cite{Shou:2014eya}. The parameters for the two sets 
are summarized in Table~\ref{tbl:WS_parameters}.

The parameterization of Shou $et~al.$ differs from the more conventional 
one of Masui $et~al.$ in two ways. First, the finite radius of the nucleon 
is taken into account. Second, rather than taking the mean radius and 
diffuseness for the deformed nucleus used in Equations~\ref{eqn:UUshape1} 
and~\ref{eqn:UUshape2} directly from electron scattering experiments, 
their values are chosen so that after averaging over all orientations of 
the axis-of-symmetry for the nucleus, the average radius and diffuseness 
match the values reported from electron scattering experiments. The result 
is that the surface diffuseness, $a$, is considerably smaller for set 2 
than for set 1, while the other parameters are similar in value.

\begin{table}[!htb]
  \caption{\label{tbl:WS_parameters} 
Parameters of the deformed Woods-Saxon distribution used in 
Equations~\ref{eqn:UUshape1} and~\ref{eqn:UUshape2}.
}
  \begin{ruledtabular} \begin{tabular}{ccccc}
&  Parameter & set 1 & set 2 & \\
  \hline
&  $R~(fm)$   & 6.81 &  6.86 & \\
&  $a~(fm) $   & 0.6 &  0.42 & \\
&  $\beta_2$   & 0.28 &  0.265 & \\
&  $\beta_4$   & 0.093 &  0 & \\
  \end{tabular} \end{ruledtabular}
\end{table}

The smaller surface diffuseness of set 2 results in a notably more compact 
nucleus.  The total inelastic \uu cross section is 8.3 barns for set 1 and 
only 7.3 barns for set 2.  The result is that the average number of binary 
collisions is 287 for set 1 and 323 for set 2.  Although set 2 appears to 
have a more consistent usage of the electron scattering data, we note that 
the neutron skin thickness of large nuclei is not probed via electron 
scattering and is thus rather unconstrained.

\begin{figure}[!htb]
    \includegraphics[width=1.0\linewidth]{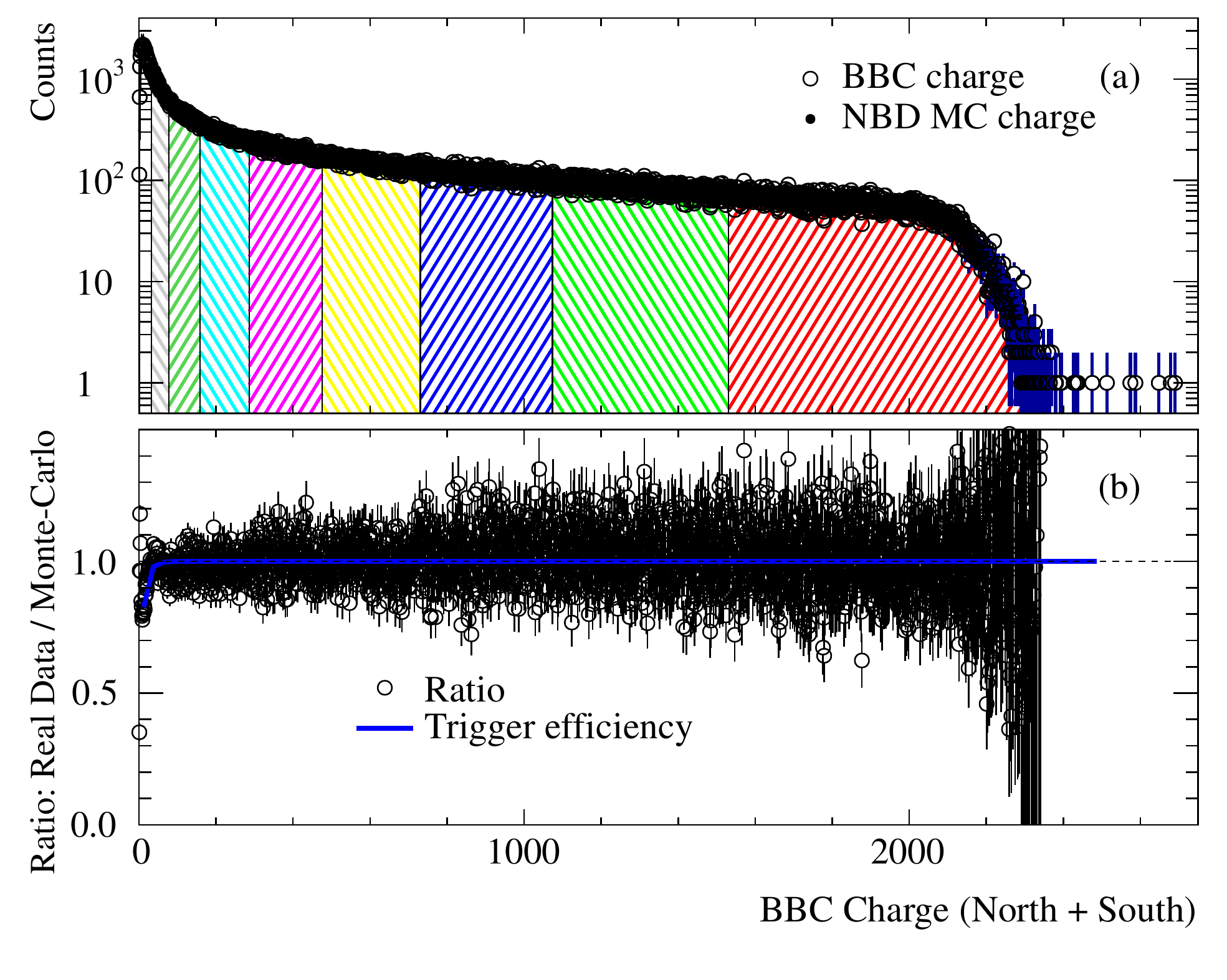}
    \caption{\label{fig:BBCdistribution} (Color online)
The BBC charge distribution for \uu collisions for the north and south 
detectors combined, compared with a Monte-Carlo calculation using a 
negative binomial distribution that is fit to the data. The colored stripes 
show the BBC charge distribution divided into 10\% wide centrality bins.
    }
\end{figure}

The Glauber model was folded with a negative-binomial distribution, 
which models the charge production in the rapidity range of the BBC. 
The parameters of the negative-binomial distribution were fit to the 
measured charge distribution from the BBC only in the signal range 
where the BBC minimum-bias trigger is known to be fully efficient. In 
the low signal range, the efficiency of the trigger was then 
determined from the ratio of the measured BBC charge distribution to 
the fitted-negative-binomial distribution. 
Figure~\ref{fig:BBCdistribution} shows the distribution of measured 
BBC charge.  Figure~\ref{fig:BBCdistribution}(a) shows the measured 
charge distribution compared with the charge distribution obtained 
from a Monte-Carlo calculation using the fitted-negative-binomial 
distribution. The efficiency of the BBC trigger was found to be 
96$\pm$3\%.  Figure~\ref{fig:BBCdistribution}(b) shows the ratio of 
data to the Monte-Carlo calculation.

Using the measured BBC charge, the events were divided into 10\% wide 
centrality bins, as illustrated in Fig.~\ref{fig:BBCdistribution}(a). 
Due to the limited statistical precision of the data sample, yields were 
extracted only for centralities from 0\%--80\%. The mean number of 
participating nucleons (\Npart) and mean number of nucleon-nucleon 
collisions (\Ncoll) for each centrality bin were found from the 
Monte-Carlo-Glauber calculation. In both cases the Glauber model used a 
nucleon-nucleon cross section of 42 mb and assumed a hard core radius of 
0.4 fm for the distribution of nucleons in the nucleus. The values of 
these parameters are summarized in Table~\ref{tbl:Glauber} for U$+$U, as 
determined using deformed Woods-Saxon parameter set 1 and 2.

\begin{table*}
  \caption{\label{tbl:Glauber} 
Centrality parameters \Npart and \Ncoll in U$+$U (this work) and Au$+$Au~\cite{Adare:2011yf} collisions, estimated using the Glauber model.
The systematic uncertainties are shown, and were estimated as described in the text. 
}
\begin{ruledtabular} \begin{tabular}{ccccccccc}
&   & \multicolumn{2}{c}{U$+$U set 1} &  \multicolumn{2}{c}{U$+$U set 2} 
& \multicolumn{2}{c}{Au$+$Au}            
\\ 
&  Centrality  & $N_{\rm part}$         & $N_{\rm coll}$               
&  $N_{\rm part}$          & $N_{\rm coll}$             
& $N_{\rm part}$  & $N_{\rm coll}$   
\\ \hline
&   0\%--10\%     & 386\,$\pm$\,5.2  & 1161\,$\pm$\,126   &  387\,$\pm$\,5.5    & 1228\,$\pm$\,142  &  326$\pm$\,3.9   & 962\,$\pm$\,97     & \\
&  10\%--20\%     & 273\,$\pm$\,6.7 & 708\,$\pm$\,69      &  274\,$\pm$\,6.4   & 770\,$\pm$\,86       &  236$\pm$\,5.5   & 609$\pm$\,60     & \\
&  20\%--30\%     & 190\,$\pm$\,6.6 & 426\,$\pm$\,42      &  192\,$\pm$\,6.7   & 473\,$\pm$\,52       &   168$\pm$\,5.8  & 378$\pm$\,37     & \\
&  30\%--40\%     & 128\,$\pm$\,6.7 & 244\,$\pm$\,24      &  130\,$\pm$\,6.5   & 277\,$\pm$\,29       &   116$\pm$\,5.8  & 224$\pm$\,23     & \\
&  40\%--50\%     &  81.9\,$\pm$\,6.4 & 130\,$\pm$\,16     &  83.0\,$\pm$\,6.2 & 149\,$\pm$\,18       &   76.2$\pm$\,5.5 & 125$\pm$\,15     & \\
&  50\%--60\%     &  47.7\,$\pm$\,5.4 &  61.7\,$\pm$\,10   &  48.5\,$\pm$\,5.3 &  71.0\,$\pm$\,11.3  &   47.1$\pm$\,4.8  & 63.9$\pm$\,9.4   & \\
&  60\%--70\%     &  24.7\,$\pm$\,4.0 &  25.8\,$\pm$\,5.3  &  25.3\,$\pm$\,3.9 &  29.3\,$\pm$\,6.0    &   26.7$\pm$\,3.7  & 29.8$\pm$\,5.4  & \\
&  70\%--80\%     &  10.9\,$\pm$\,2.3 &  9.2\,$\pm$\,2.4    &  11.2\,$\pm$\,2.4 &  10.2\,$\pm$\,2.8    &   13.7$\pm$\,2.5  & 12.6$\pm$\,2.8 & \\ 
  \end{tabular} \end{ruledtabular}
\end{table*}

The systematic uncertainties for the mean \Npart and \Ncoll values in each 
centrality bin were estimated by varying the parameters of the Glauber 
model within reasonable limits. The nucleon-nucleon cross section was 
varied from 42 mb down to 39 mb and up to 45 mb. For deformed Woods-Saxon 
parameter set 1 the radius $R$ and diffuseness parameter $a$ were varied 
down to $R=6.50$ fm and $a=0.594$ fm and up to $R=6.92$ fm and $a=0.617$ 
fm. These variations were chosen to match the percentage variations used 
for the \auau case, as were the variations used in the evaluation of 
systematics for set 2. In addition to these variations of the Glauber 
model parameters, the calculation was run without using a hard core for 
the nucleon distribution, and the trigger efficiency was varied within its 
uncertainty of 3\%. For set 1, the uncertainty in the U deformation 
parameters has been estimated to be approximately 3\% for the dipole 
component $\beta_2$ and approximately 50\% for the small contribution of 
the quadrupole component $\beta_4$~\cite{Milner:1977zz}. Varying $\beta_2$ 
and $\beta_4$ by these amounts resulted in an insignificant contribution 
to the systematic uncertainty for \Npart and \Ncoll. The systematic 
uncertainties are shown in Table~\ref{tbl:Glauber}.

The \Npart values obtained from the two deformed Woods-Saxon parameter 
sets are essentially indistinguishable. However the differences between 
the \Ncoll values obtained from the two parameter sets are, at some 
centralities, outside the uncertainties on the \Ncoll values. This is due 
primarily to the large difference in the diffuseness values for the two 
sets. As noted earlier, the smaller diffuseness for set 2 (0.43 fm vs 0.6 
fm for set 1) combined with a similar mean radius results in larger \Ncoll 
values at all centralities because it makes the nucleus more compact.

Because the \Ncoll values are important in the interpretation of the U$+$U 
data, and are directly involved in the calculation of the \raa, we have 
chosen to consider the \Ncoll values from both sets 1 and 2 in the 
remainder of the paper.

	\subsection{Muon-Track and Pair Reconstruction}

Muon candidates are charged particle tracks which penetrate all layers of the MuID. 
These MuID tracks are matched to MuTr tracks, which provide the 
momentum measurement. Although the requirement that tracks pass 
through the whole spectrometer arm reduces significantly the hadron 
contribution, a small percentage of charged hadrons can travel through 
without interacting ($\sim$0.1\%), and these are a source of 
background. Some muons produced from charged hadron decays in front of 
the MuTr are also reconstructed, and form a background of real muons 
in the spectrometer arms. Various offline analysis selection criteria 
are used to enhance the sample of good muon candidate tracks. There is 
a cut on the single track $\chi^{2}$, and also on the difference in 
position and angle between the extrapolated MuTr and MuID parts of the 
candidate track.  During the dimuon reconstruction, the selected track 
pair is fit with the event $z$-vertex position, and a second 
$\chi^{2}$ cut is applied for the track pair fit.

	\subsection{$\mu^{+}$+$\mu^{-}$ Analysis}\label{sec:Analysis}

The invariant mass distribution is formed by combining all pairs of 
oppositely charged muon tracks. There is a significant combinatorial 
background under the \jpsi peak that is formed by single muons or 
mis-identified hadrons that randomly combine to form a pair. There is also 
a continuum due to correlated muons from semi-leptonic decays of open 
charm and bottom, and correlated muons from the Drell-Yan process. The 
combinatorial background may be estimated by event mixing, in which tracks 
with opposite charges from different events (of similar $z$-vertex and 
centrality) are combined randomly - see Ref.~\cite{Adare:2011yf} for more 
details. By construction, such mixing destroys any real muon correlations, 
and so the mixed event background does not reproduce the correlated 
background from physics sources.

\begin{figure*}[!htb]
   \includegraphics[width=0.47\linewidth]{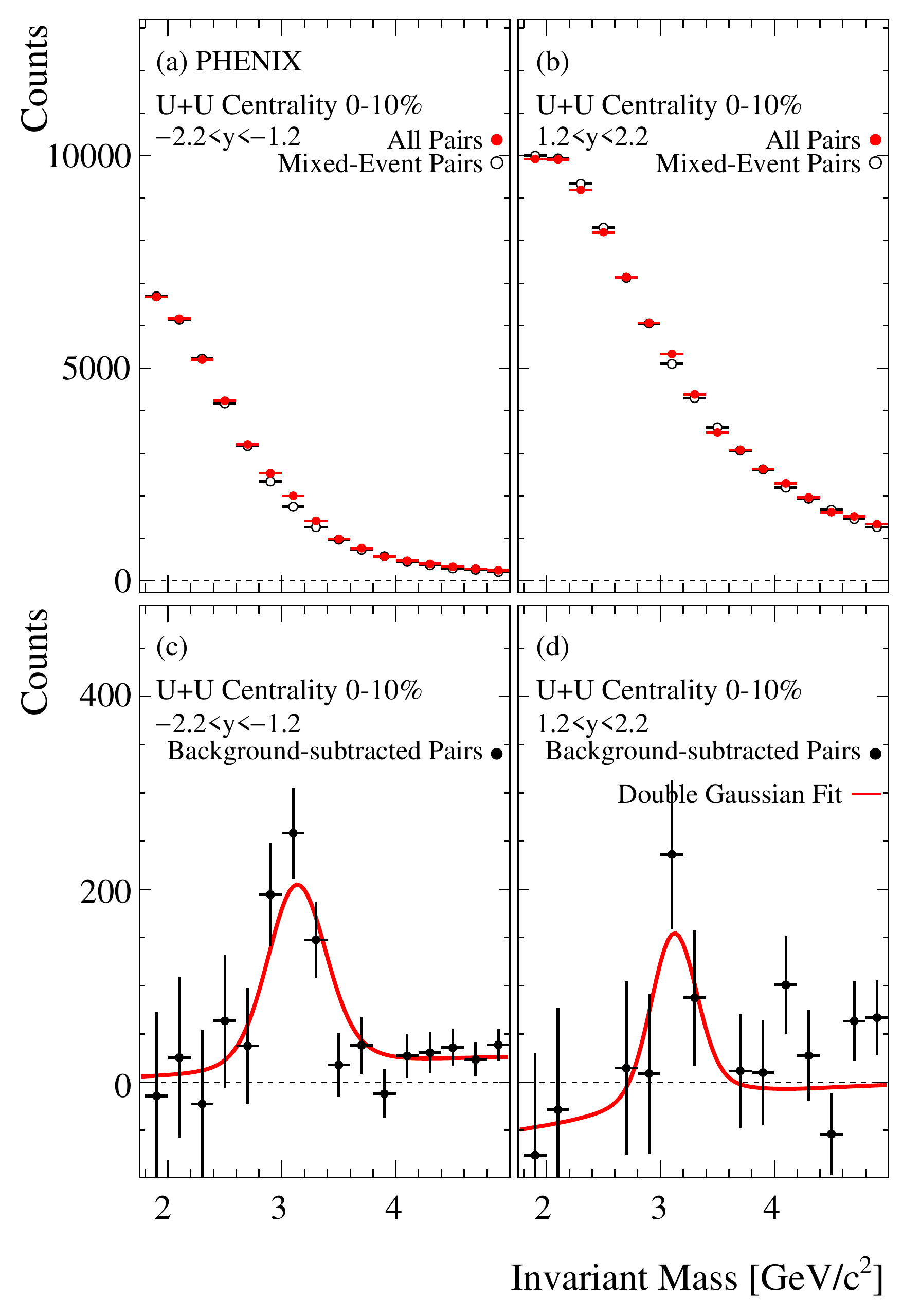}
    \includegraphics[width=0.47\linewidth]{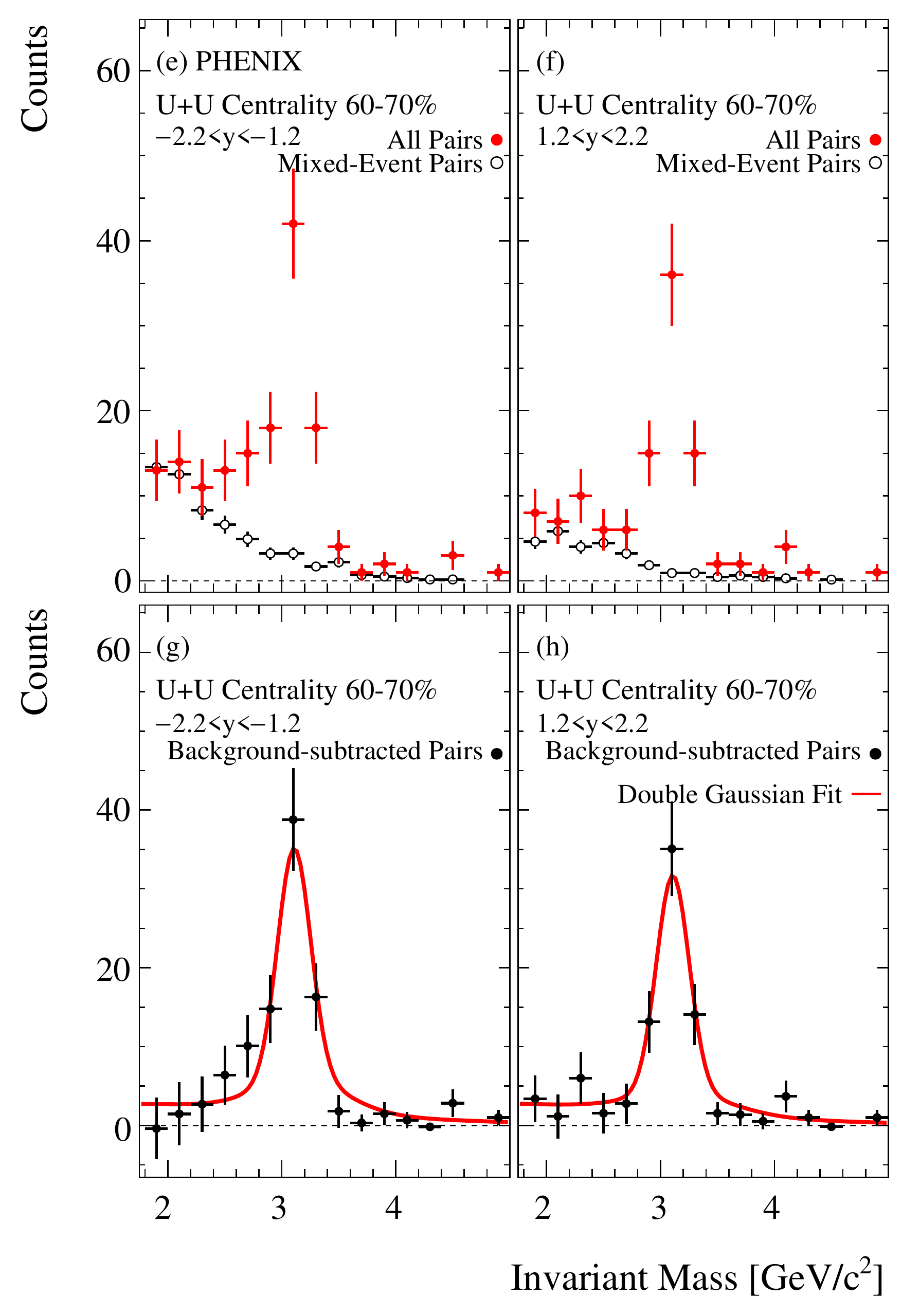}
\caption{\label{fig:RawCounts}  (Color online)
Dimuon invariant mass spectra at forward and backward rapidity measured 
in the 0\%--10\% most central collisions (panels (a)--(d) )and in 60\%--70\% mid-peripheral 
collisions (panels (e)--(h)), integrated over the full \pt range.  Panels (a), (b), (e), and 
(f) show the invariant mass distribution, reconstructed from (solid 
symbols) same-event opposite charge-sign pairs and (open symbols) 
mixed-event pairs in \uu collisions.  Panels (c), (d), (g), and (h)
show the combinatorial background subtracted mass spectra.  
The solid line is a fit to the data using a double Gaussian line shape 
plus an exponential background, as described in the text. 
}
\end{figure*}

Figure~\ref{fig:RawCounts}(a)--(h) show the dimuon spectra from the 
south and north spectrometer arms for two example centrality bins and 
the combinatorial-background-subtracted mass spectra. The mixed-event 
combinatorial background is normalized to the real data using 
yields of muon pairs having the same charge sign found in both the mixed-pairs and real-pairs data 
samples in a range close to the \jpsi mass peak region, 
2.6\mylessthan$M$\mylessthan3.6\,GeV/$c^{2}$. The procedure is similar 
to that described in Ref.~\cite{Adare:2011vq}.  Because the mass 
resolution precludes the separation of the \jpsi and $\psi'$ peaks in 
this analysis, we present only \jpsi results in this paper.

The \jpsi yield was obtained using a fit in the mass range 1.7 to 5 
\,GeV/$c^{2}$. The fitting function included the mixed event background, a 
signal line shape, and an exponential. The signal line shape for the \jpsi 
was a double Gaussian line shape modified by the mass dependence of the muon arm acceptance. 
The exponential was 
used to account for the correlated background that is not described by the 
mixed-event background, and to compensate for any systematic effects from 
under or over-subtracting the large combinatorial background. Because of 
the limited statistical sample for the \uu measurement, the form of the 
signal line shape was based on studies of the mass spectrum in \pp 
collisions, with input from studies performed for the \cuau 
system~\cite{Aidala:2014bqx} where the \jpsi mass width was found to 
increase linearly with increasing multiplicity in the muon arms. The 
functional form of the increase of the width with collision centrality was 
found independently for each muon arm from \cuau collision data, taken in 
the same RHIC data-taking period.  This was then extrapolated where 
necessary to the higher multiplicity for the \uu system.  As a cross 
check, a similar increase of the widths with multiplicity was observed in 
a GEANT~\cite{Brun:GEANT} Monte-Carlo calculation with simulated 
{\sc pythia}~\cite{Sjostrand:PYTHIA} \jpsi events embedded into real \uu 
data events. The mass width varies from (peripheral to 
central) 0.14 to 0.2~GeV/$c^2$ for the south arm, and 
from (peripheral to central) 0.14 to 0.24~GeV/$c^2$ in the north arm.

Because the width was fixed at each centrality by this procedure, the \jpsi 
yield was the only free parameter in the line shape.  The fit is shown in 
Fig.~\ref{fig:RawCounts}(c), (d), (g), and (h).  The systematic uncertainty 
on the yield from the fit was estimated by varying the mass fit range, 
varying the relative normalization of the signal to the mixed event 
combinatorial background by $\pm$2\%, and also extracting the yield by 
using a like-sign combinatorial background in the fit instead of 
the mixed-event background.  Additional systematic checks were made by 
comparing the 
yields to those obtained from the raw signal count after exponential 
background subtraction, and to those obtained from other fit functions. 
The latter included allowing the Gaussian widths of the \jpsi peak to be 
free in the fit function, where it was found that the yields agreed within 
the statistical uncertainty. The systematic uncertainty from the fit was 
larger in the north arm than in the south arm. For both arms, the fit 
systematic uncertainties were estimated to range from 0.8\% for peripheral 
events to 10.9\% for central events.

\begin{figure}[!hbt]
    \includegraphics[width=1.0\linewidth]{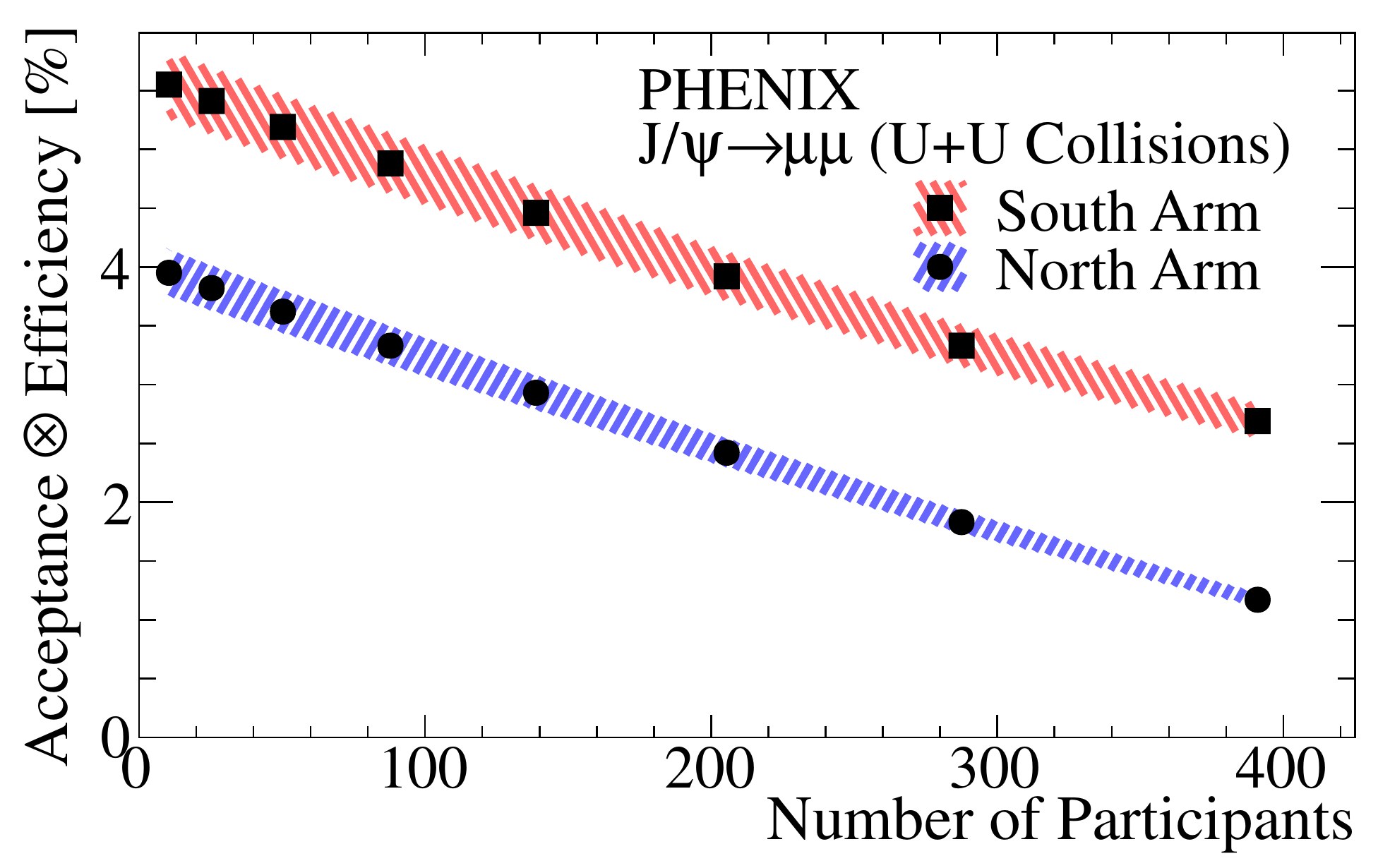}
    \caption{\label{fig:Efficiency} (Color online)
Acceptance $\otimes$ efficiency as a function of collision centrality 
($\Npart$) for (squares) south arm, \ybackward and (circles) north arm, 
\yforward.  The bands represent the uncertainty due the limited statistical precision of the embedding simulations.
    }
\end{figure}

	\subsection{Efficiency and Corrections}
	\label{efficiency}

The \jpsi detection efficiency was estimated in a two-step process. First, 
the efficiency to detect hits in each plane of the MuID was estimated by 
finding all roads made by charged particles through the MuID, but ignoring 
hits in the plane of interest.  The efficiency for the plane of interest 
was then estimated from the number of roads with associated hits in that 
plane compared with all roads ignoring hits in that plane. Because the 
MuID efficiency decreases with increasing luminosity, the final efficiency 
used was the luminosity weighted average over all runs used in this 
analysis. The MuID efficiency is included in the calculation of the 
acceptance\,$\otimes$\,reconstruction efficiency described below.

The full acceptance\,$\otimes$\,efficiency (acceptance convoluted with efficiency) is estimated by embedding 
{\sc pythia} \jpsi$\rightarrow\mumu$ decays -- fully simulated via a 
{\sc geant} description of the PHENIX detector -- into a reference data 
sample. The data plus simulation is then reconstructed with the same 
analysis as the real data and the final acceptance\,$\otimes$\,efficiency 
is evaluated, normalized by the number of simulated 
\jpsi$\rightarrow\mumu$ decays over the same rapidity range.

\begin{table}[!htb]
  \caption{\label{tbl:Uncertainties} Estimated systematic uncertainties.}
  \begin{ruledtabular} \begin{tabular}{ccc}
  Source & Uncertainty (\%) & Type \\
  \hline
  \jpsi signal extraction   & $\pm$0.8--10.9 & A \\
  input \jpsi \pt distributions & $\pm$4.0       & B \\
  detector acceptance & $\pm$5.0       & B \\%
  reconstruction and trigger efficiency  & $\pm$5.0      & B \\
  run-to-run efficiency variation & $\pm$2.8       & B \\
  Glauber (\Ncoll)    & $\pm$10.8--33.0    & B \\
  \pp reference    & $\pm$7.1    & C \\
  \pp reference energy scale    & $\pm$3.6  & C \\
  \end{tabular} \end{ruledtabular}
\end{table}

The acceptance $\otimes$ efficiency for \jpsi in \uu collisions is shown 
in Fig.~\ref{fig:Efficiency}.  As observed for other collision systems, 
the north arm has lower efficiency than the south, and the efficiency is 
strongly centrality dependent. These factors are reflected in the final 
data yields, where the statistical uncertainties are largest for the north 
arm, and increase with increasing centrality.

Between the measurement of the \pp data~\cite{Adare:2010fn} used as the 
reference for the \uu nuclear modification (see section~\ref{raa}) and the 
\uu data, additional absorber material was added in front of the muon 
arms. The added absorber increases the minimum energy required for a muon 
to penetrate to the last gap of the MuID detector, and this reduces the 
\jpsi acceptance near $y$ = 1.2 at low \pt. Because a realistic rapidity 
shape (from {\sc pythia}) is used when calculating the rapidity integrated 
acceptance\,$\otimes$\,efficiency, it should be correct for both the \uu 
and \pp measurements. We note that the systematic uncertainty for the \uu 
acceptance\,$\otimes$\,efficiency includes an uncertainty due to the 
possible deviation of the \uu rapidity shape from that given by {\sc 
pythia}, as discussed in section~\ref{systematic}.

	\subsection{Systematic Uncertainties}
	\label{systematic}

The systematic uncertainties are divided into three groups: Type A - 
point-to-point uncorrelated uncertainties, Type B - point-to-point 
correlated uncertainties, and Type C - global scale uncertainties, which 
are summarized in Table~\ref{tbl:Uncertainties}. The signal extraction 
systematic uncertainty due to the fitting procedure, discussed in 
Section~\ref{sec:Analysis}, is treated as a Type A uncertainty. The 
uncertainty arising from the assumptions about the input \jpsi momentum 
and rapidity distributions used in the {\sc pythia} calculations were 
previously studied in~\cite{Adare:2012wf} and were found to be 
$\sim$4.0\%. The uncertainty in the detector acceptance is estimated to be 
$\sim$5.0\%. An overall uncertainty on the reconstruction and trigger 
efficiency obtained from embedding {\sc pythia} events in real data is 
estimated to be $\sim$5.0\%. Small run-to-run variations in the MuID and 
MuTr efficiency were studied in~\cite{Aidala:2014bqx} and found to be 
2.8\%.  The uncertainty in the mean \Ncoll values for the centrality bins 
was studied as described in section~\ref{sec:centrality}, and found to be 
10.8-33.0\% (see Table~\ref{tbl:Glauber}). 
The measured \pp reference invariant yield contributes a systematic uncertainty of 7.1\%. This is smaller than the 
\pp global systematic uncertainty because both the BBC trigger efficiency for \pp and the \Ncoll 
estimate used in the \raa depend on the assumed nucleon-nucleon cross section,
and so their systematic uncertainties cancel in part when forming \raa.
An additional small systematic uncertainty  
is assigned to the \pp reference for the nuclear-modification factor because the \pp cross 
section, measured at \sqrts = 200~GeV, had to be extrapolated to \sqrts = 
193~GeV to obtain the reference cross section for the \uu data set. That 
systematic uncertainty was taken conservatively to be equal to the 
correction, 3.6\%, and when it is added in quadrature with the 7.1\% from the \pp reference measurement, 
the overall global uncertainty due to the \pp reference becomes 8.1\%.

		\section{Results}

\begin{figure}[!htb]
    \includegraphics[width=1.0\linewidth]{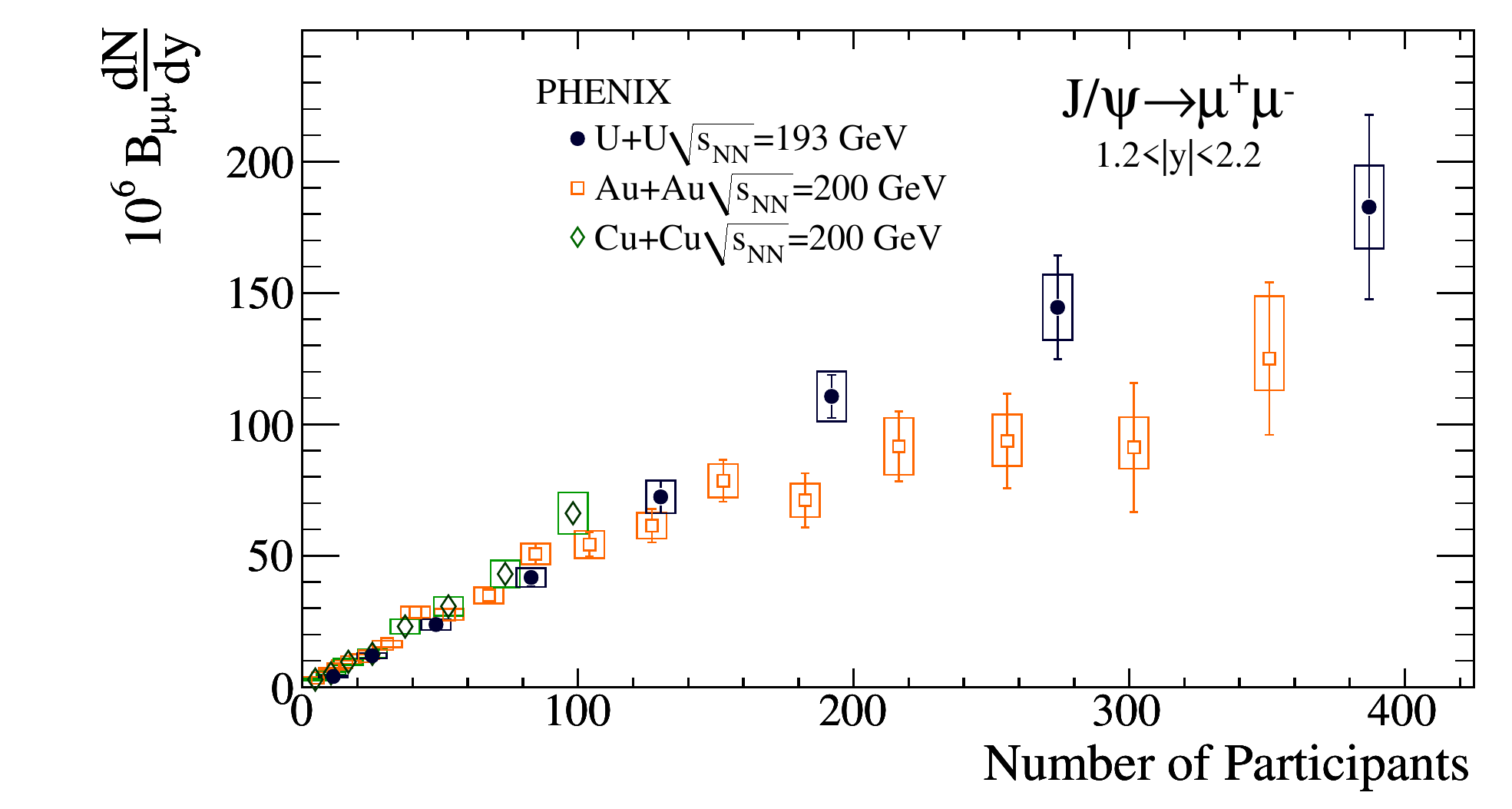}	
  \hspace{0.025\textwidth}
    \caption{\label{fig:Yield} (Color online) 
Invariant yield of the \jpsi at forward rapidity measured as a function of 
collision centrality ($\Npart$) for (closed circles) \uu, (open squares)
\auau~\protect\cite{Adare:2011yf}, and (open diamonds) 
\cucu~\protect\cite{Adare:2008sh}.
    }
\end{figure}

The \jpsi yields measured in the two muon arms agree well, within 
statistical uncertainties. Therefore the \uu results presented here are 
averaged over both forward and backward rapidity.

\subsection{Yield and $R_{AA}$ versus \Npart}
\label{raa}
The measured invariant yield is obtained from  Eq.~\ref{eqn:InvYield}:

\begin{equation}
\label{eqn:InvYield}
\ensuremath{B_{\mu\mu}\frac{dN}{dy} = \frac{1}{N_{\rm event}}\frac{N^{\jpsi}_{measured}}{{\Delta}yA\epsilon},}
\end{equation} 

\noindent
where $B$ is the branching fraction for \jpsi decay to dimuons, 
$N^{\jpsi}_{measured}$ is the measured number of \jpsi integrated over all 
\pt, $N_{\rm event}$ is the number of minimum-bias events analyzed, 
$A\epsilon$ is the acceptance\,$\otimes$\,efficiency, and ${\Delta}y$ is 
the rapidity range used when calculating the acceptance of the muon arms.

The invariant yield for \jpsi production in \uu collisions at 
$\sqrt{s_{NN}}$\,=\,193\,GeV is shown versus \Npart in 
Fig.~\ref{fig:Yield}, where the vertical error bars represent the 
statistical uncertainty plus Type A systematic uncertainty for the yield 
extraction, added in quadrature. The boxes represent the Type B systematic
uncertainties, summed in quadrature. In addition to \uu, the largest 
system measured at RHIC, the \Npart dependence of the invariant yield is 
shown for two other symmetric systems, \auau~\cite{Adare:2011yf} and 
\cucu~\cite{Adare:2008sh} at \sqrtsn\,=\,200\,GeV. The yields below 
$N_{\rm part}$\,$\approx$\,150 are similar for all three systems at the 
same \Npart.  However for $N_{\rm part}$\,$\gtrsim$\,200 the yield for \uu 
is larger than that for \auau.

The departure from unity of the nuclear-modification factor,
\begin{equation}
\label{eqn:RAA}
\ensuremath{R_{AA} = \frac{1}{\langle \Ncoll 
\rangle}\frac{dN^{AA}/dy}{dN^{pp}/dy}},
\end{equation} 
quantifies the modification of the \Ncoll normalized invariant yield in 
heavy ion collisions relative to the invariant yield in \pp collisions. 
The reference \pp invariant yield was obtained from~\cite{Adare:2010fn}. 
Because the \pp data were measured at \sqrts=200~GeV a scale factor 
of 0.964, determined from \pp~{\sc pythia} simulations at 
\sqrts=193~GeV and \sqrts=200~GeV, was applied to the \pp 
invariant yield to account for the difference in \jpsi cross section.

\begin{figure}[!htb]
    \includegraphics[width=1.0\linewidth]{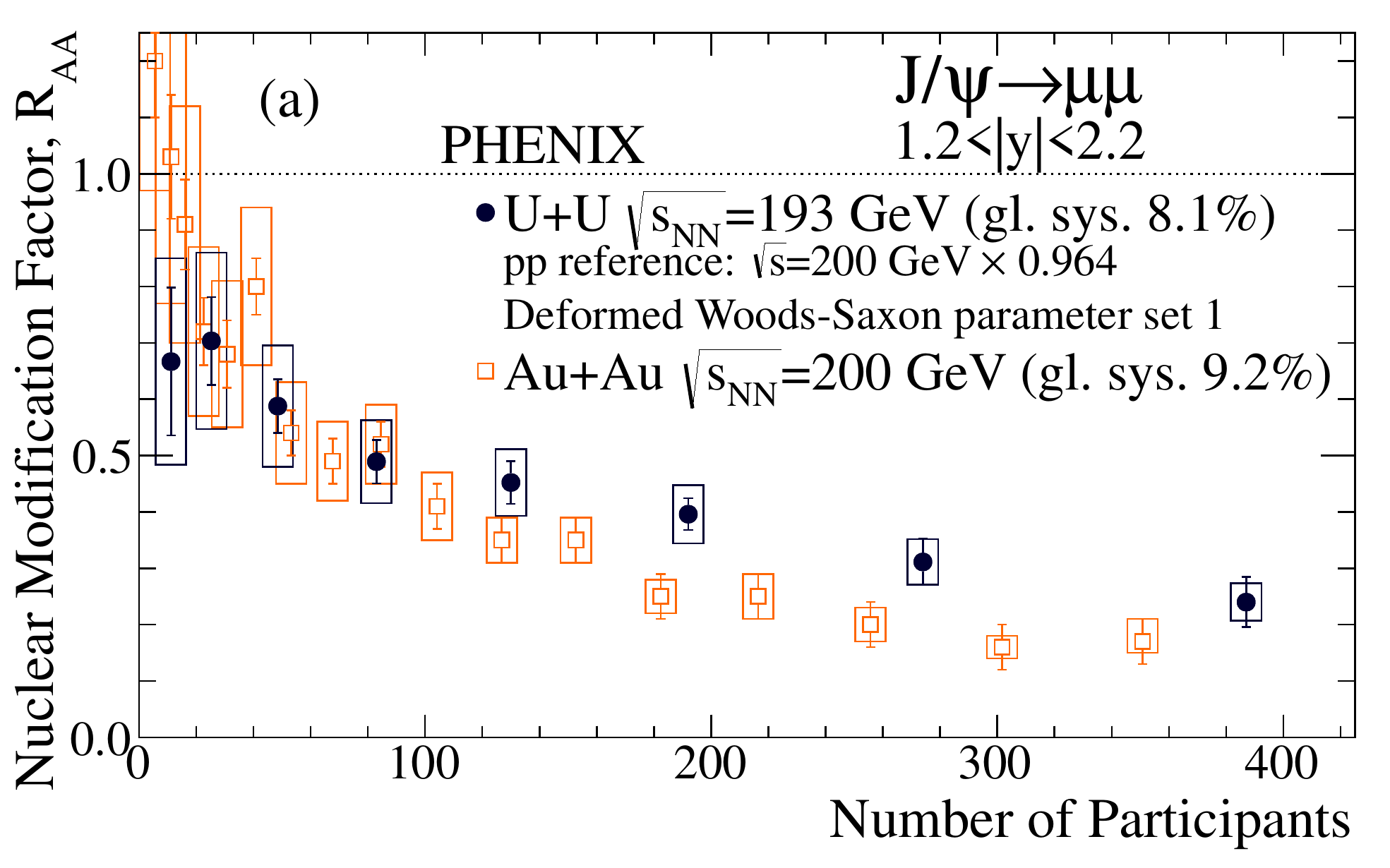}
    \includegraphics[width=1.0\linewidth]{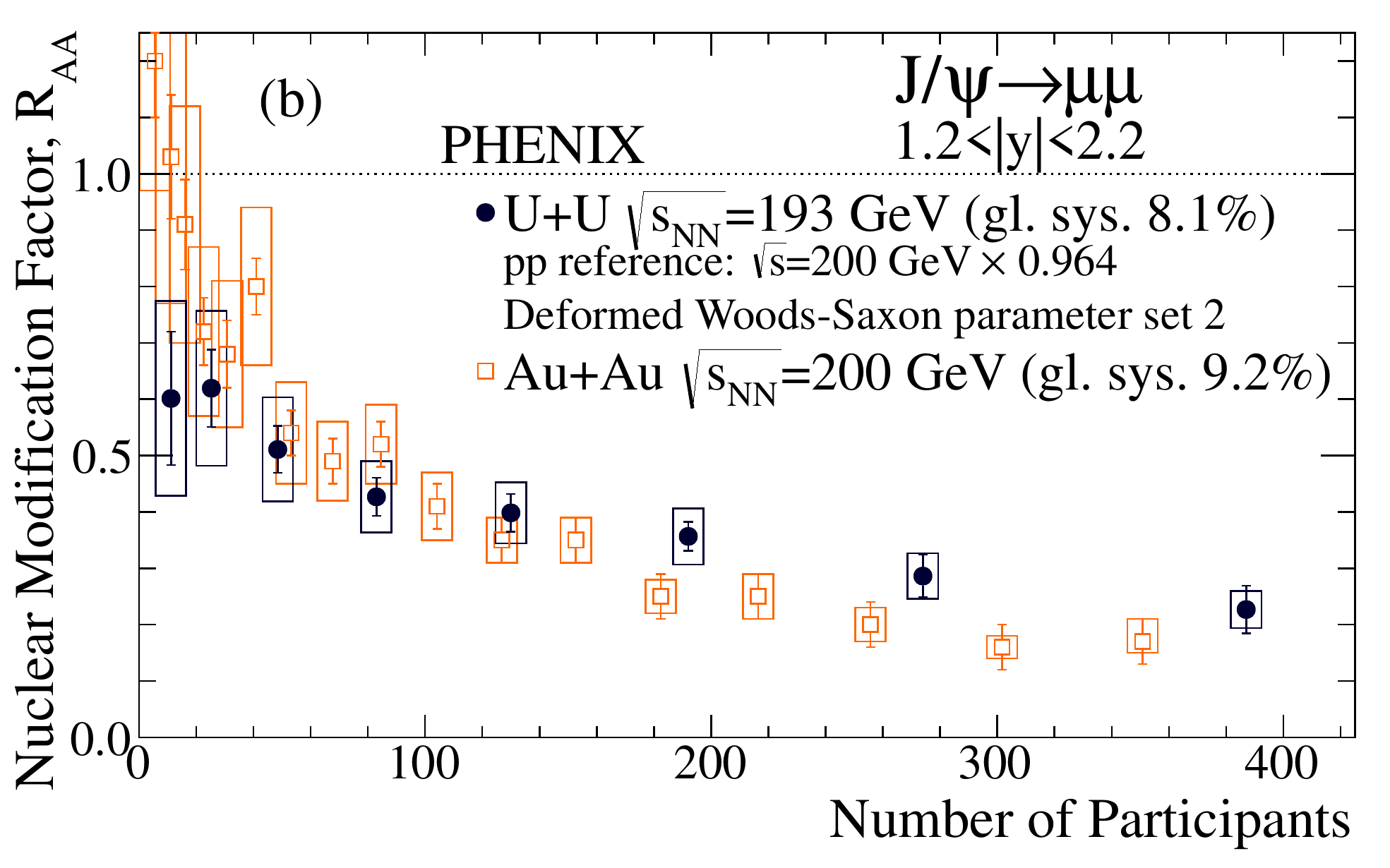}
    \caption{\label{fig:RAA} (Color online) 
The nuclear-modification factor, \raa, measured as a function of collision 
centrality ($\Npart$) for (closed circles) $\jpsi$ at forward rapidity in 
\uu collisions, for which the \pp reference data, measured 
at \sqrts=200~GeV, has been scaled down by a factor 0.964 to account 
for the difference in \jpsi cross section between \sqrts=200 and 193 
GeV.  (open squares) Au$+$Au data~\protect\cite{Adare:2011yf}.
(a) The U$+$U \raa calculated using Woods-Saxon parameter set 1~\cite{Masui:2009qk} in the U$+$U Glauber 
model calculation. (b) The U$+$U \raa 
calculated using parameter set 2~\cite{Shou:2014eya}.
    }
\end{figure}

As discussed in section~\ref{sec:centrality}, the deformed Woods-Saxon 
parameter sets 1 and 2 are derived using different assumptions, resulting 
in substantially different values of the surface diffuseness parameter, 
$a$. The authors of~\cite{Shou:2014eya} argue that their approach in 
producing parameter set 2 corrects deficiencies in the conventional 
method, and is the correct one. In any case, both assumptions cannot be correct,
so the differences in the deformed Woods-Saxon parameters between the two sets do 
not represent a systematic uncertainty on the \raa, because one set is correct (within its 
uncertainties) and the other is not. We have chosen to present the \raa 
calculated using both sets, so that the effect of using a conventional 
description and the description of reference~\cite{Shou:2014eya} of the 
deformed U nucleus can be compared.

The nuclear-modification factor for \uu collisions is shown as a function 
of \Npart in Fig.~\ref{fig:RAA}. Fig.~\ref{fig:RAA} (a) shows the \raa 
calculated using \Ncoll values from deformed Woods-Saxon parameter set 1. 
Fig.~\ref{fig:RAA} (b) shows the \raa calculated using those from 
parameter set 2. The vertical bars represent the combined statistical and 
Type A systematic uncertainties and the boxes represent the Type B 
uncertainties. The Type C (global) uncertainties are listed in the legend.
The overall global uncertainty (Type C) is 8.1\%. The 
modification for \auau collisions~\cite{Adare:2011yf} is shown for 
comparison. The \raa values for \uu collisions are provided in 
Table~\ref{tbl:RAA}.

The measured \raa for the \uu collision system is similar to that for the 
\auau system, although the \uu data for the most central collisions show 
less modification than the \auau data.

\begin{table*}
  \caption{\label{tbl:RAA} 
The nuclear-modification factor ($\raa$) for \uu collisions
(averaged over forward (1.2\mylessthan$y$\mylessthan2.2) and backward 
(-2.2\mylessthan$y$\mylessthan-1.2) rapidity), as a function of centrality, 
derived from deformed Woods-Saxon parameter sets 1 and 2 (see text for details). 
The first and second uncertainties listed represent Type-A (statistical 
uncertainties plus point-to-point systematic uncertainties from the yield
extraction) and Type-B uncertainties, respectively (see text for definitions).
There is a Type-C (global) systematic uncertainty of 8.1\%.
    }
  \begin{ruledtabular} \begin{tabular}{ccccccccc}
&  Centrality & $R_{UU}$ (set 1)  & $R_{UU}$ (set 2)   \\
\hline
&  0\%--10\%  & 0.240 $\pm$ 0.045 $\pm$ 0.031 &  0.225 $\pm$ 0.042 $\pm$ 0.031  \\  
& 10\%--20\%  & 0.314 $\pm$ 0.041 $\pm$ 0.035 &  0.285 $\pm$ 0.038 $\pm$ 0.035 \\
& 20\%--30\%  & 0.400 $\pm$ 0.028 $\pm$ 0.044 &  0.349 $\pm$ 0.025 $\pm$ 0.044 \\
& 30\%--40\%  & 0.463 $\pm$ 0.039 $\pm$ 0.053 &  0.390 $\pm$ 0.033 $\pm$ 0.053 \\
& 40\%--50\%  & 0.513 $\pm$ 0.040 $\pm$ 0.070 &  0.410 $\pm$ 0.032 $\pm$ 0.070 \\
& 50\%--60\%  & 0.604 $\pm$ 0.049 $\pm$ 0.109 &  0.477 $\pm$ 0.039 $\pm$ 0.109 \\
& 60\%--70\%  & 0.698 $\pm$ 0.077 $\pm$ 0.165 &  0.567 $\pm$ 0.063 $\pm$ 0.165 \\
& 70\%--80\%  & 0.646 $\pm$ 0.137 $\pm$ 0.241 &  0.529 $\pm$ 0.104 $\pm$ 0.241 \\
  \end{tabular} \end{ruledtabular}
\end{table*}

	\subsection{Yield and $R_{AA}$ versus collision centrality}

Possible effects which may modify \jpsi~production in \uu collisions with 
respect to \auau collisions were discussed in~\cite{Kikola:2011zz}. The 
expected higher energy density in \uu compared to \auau (15\%--20\%) 
should lead to a stronger suppression due to color screening.  On the 
other hand, for a given centrality, a larger \Ncoll~value in \uu~(compared 
to \auau) should increase charm production by $c\bar{c}$ coalescence.  
Cold nuclear matter effects due to shadowing are expected to be similar in 
both systems.

\begin{figure}[!htb]
    \centering
    \includegraphics[width=1.0\linewidth]{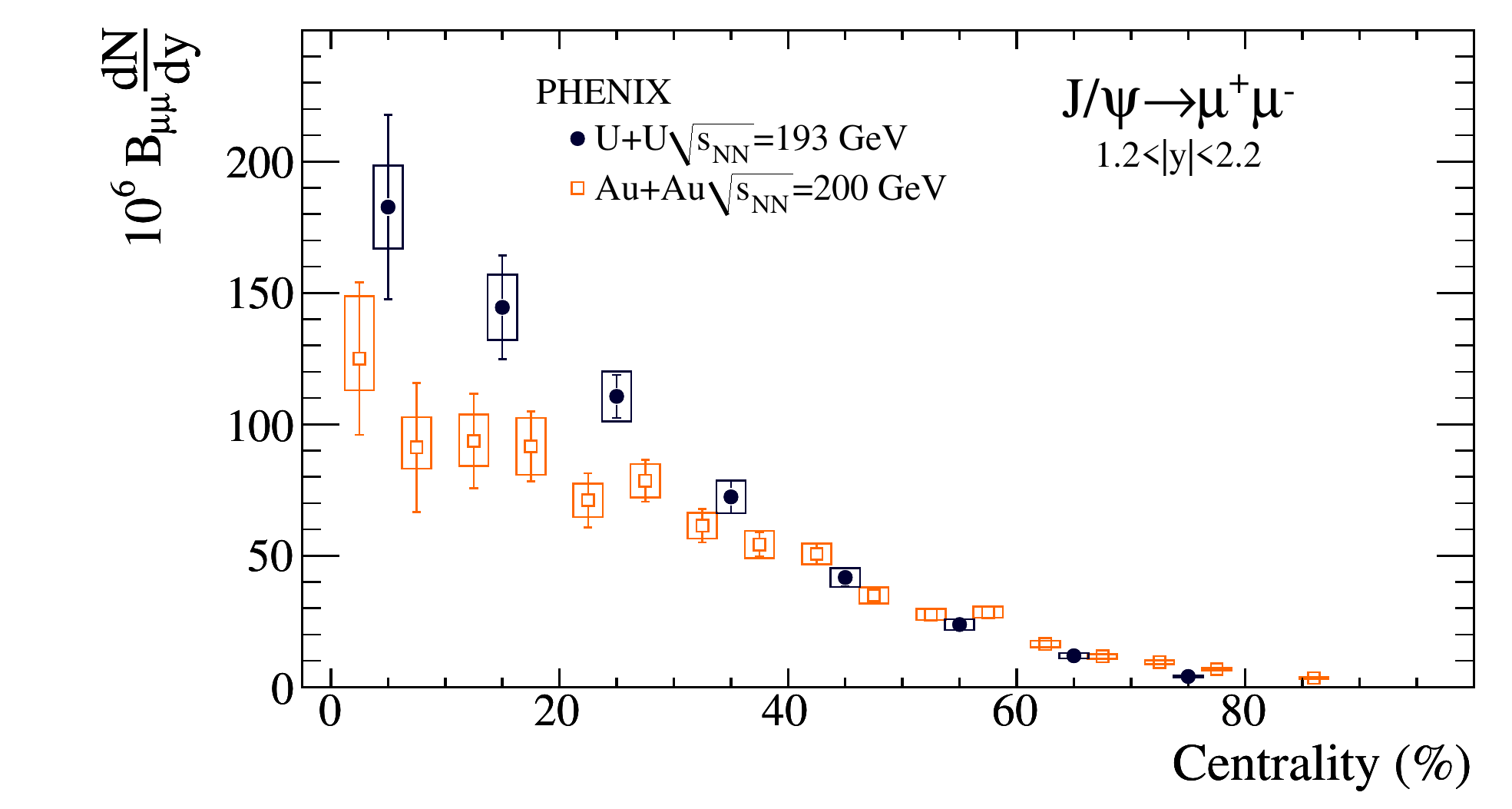}
\caption{\label{fig:Yield_XSect} (Color online) 
Invariant yield measured as a function of collision centrality for
$\jpsi$ at forward rapidity for (closed circles) \uu and  
(open squares) \auau~\protect\cite{Adare:2011yf} collisions.
}
\end{figure}

As suggested in~\cite{Kikola:2011zz}, we define for a given centrality 
class the relative nuclear-modification factor

\begin{equation}
\label{eqn:RUUAuAu}
\ensuremath{R^{UU}_{\rm AuAu} = \frac{dN^{UU}/dy}{dN^{\rm AuAu}/dy}} \bigg(\frac{N_{\rm coll}^{\rm AuAu}}{0.964 \times N_{\rm coll}^{UU}}\bigg)^2,
\end{equation} 

\begin{figure}[!htb]
    \includegraphics[width=1.0\linewidth]{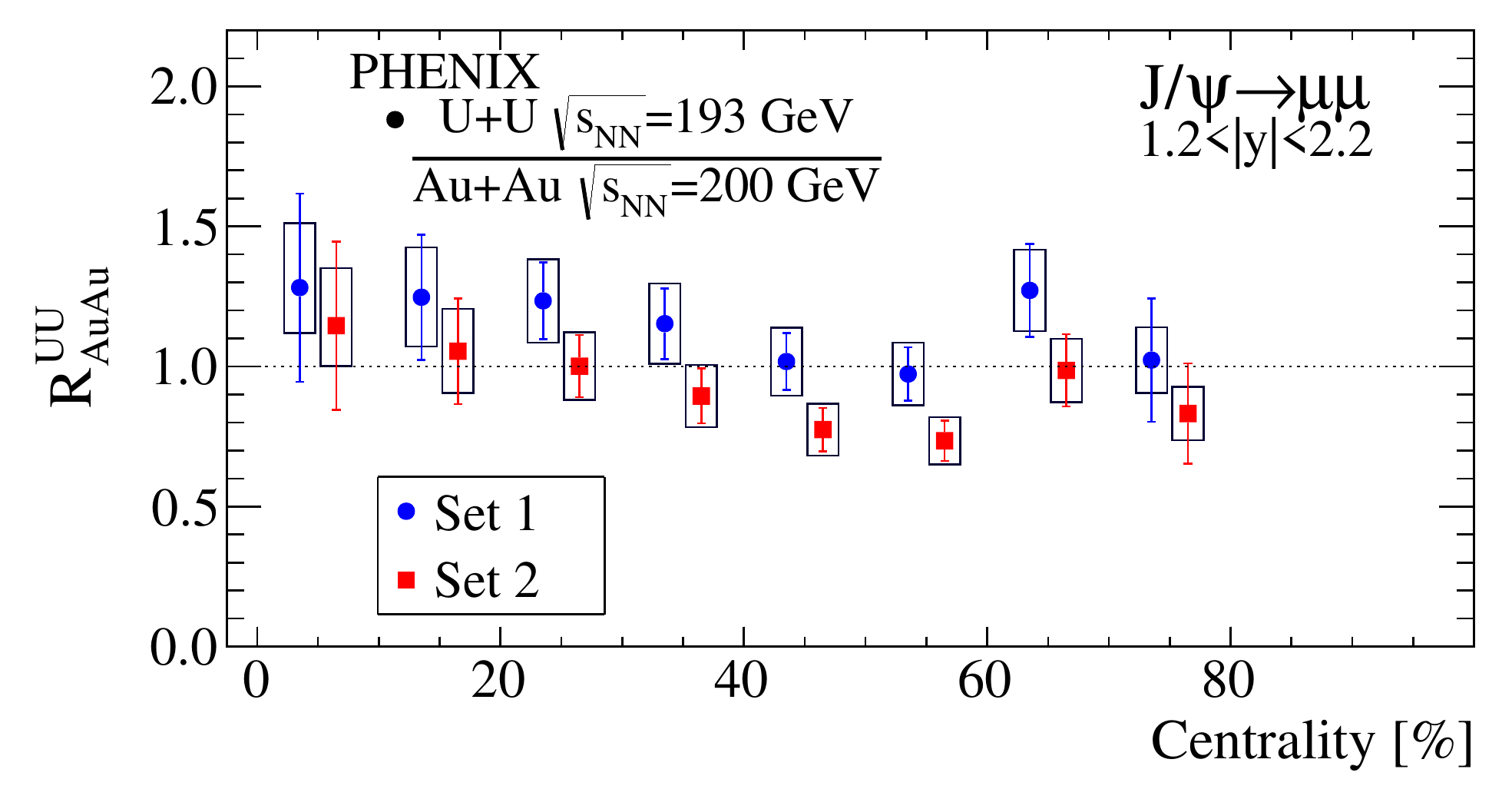}		
    \caption{\label{fig:Ratio_RAAN2_XSect} (Color online) 
The relative nuclear-modification factor for \uu and \auau 
(Equation~\ref{eqn:RUUAuAu}) as a function of collision centrality.  
Values were calculated using \Ncoll values obtained from the U$+$U Glauber model calculation using  
Woods-Saxon parameter set 1~\cite{Masui:2009qk} and set 2~\cite{Shou:2014eya}. The 
values for set 1 and set 2 are slightly offset in centrality for clarity.
}
\end{figure}

\noindent
where $N^{UU}_{\rm coll}$ is the mean value of \Ncoll in the given centrality 
class (e.g. 10\%--20\%) in \uu collisions, and $N^{\rm AuAu}_{\rm coll}$ is the 
corresponding value for \auau collisions in the same centrality class. The 
\Ncollsq ratio in Equation~\ref{eqn:RUUAuAu} is intended to account for 
the expected scaling of the \jpsi cross section with \Ncollsq in the case 
of $c\bar{c}$ coalescence. This assumes that the coalescence yield depends on the number of 
charm quarks squared, and thus on the number of binary collisions squared.
However to get the correct scaling in this 
case, the values of \Ncoll for \uu have to be modified by a factor 0.964 
because the charm production cross section in \pp collisions at \sqrts\,=\,193~\,GeV 
is smaller than that at \sqrts\,=\,200~\,GeV by that factor.

If the production of \jpsi in central collisions was entirely due to 
$c\bar{c}$ coalescence, and if cold nuclear matter effects were the same 
for both systems, the relative nuclear modification would be expected to 
be $1$.  
The variable has the advantage that it is a direct ratio of \uu and \auau 
invariant yields, eliminating the systematic uncertainties associated with 
the \pp reference when forming the \raa. The ratio of \Ncollsq values for 
\auau and \uu also appears in $R^{UU}_{\rm AuAu}$. But the systematic 
uncertainty estimation for \uu (see section~\ref{sec:centrality}) was 
carried out in such a way that the systematic uncertainties in \Ncoll are 
highly correlated for \uu and \auau, and they mostly cancel when taking 
the ratio of \Ncoll values. There is still, however, the large difference 
in the diffuseness parameter, $a$, coming from the difference between the 
models used to obtain parameter sets 1 and 2. We deal with that by 
calculating $R^{UU}_{\rm AuAu}$ for both models.

The invariant yield is shown plotted versus centrality class for \uu and 
\auau in Figure~\ref{fig:Yield_XSect}. To form the relative nuclear 
modification of Equation~\ref{eqn:RUUAuAu}, some rebinning of the \auau 
data is required. The rebinned invariant yields and their ratios are 
summarized in Table~\ref{tbl:BdNdy}.  Because the \uu and \auau data were 
taken in different years, and there were differences in the muon arm 
absorber thickness due to an upgrade, all of the systematic uncertainties 
on the invariant yields are propagated in the ratio.

 \begin{figure}[!htb]
    \includegraphics[width=1.0\linewidth]{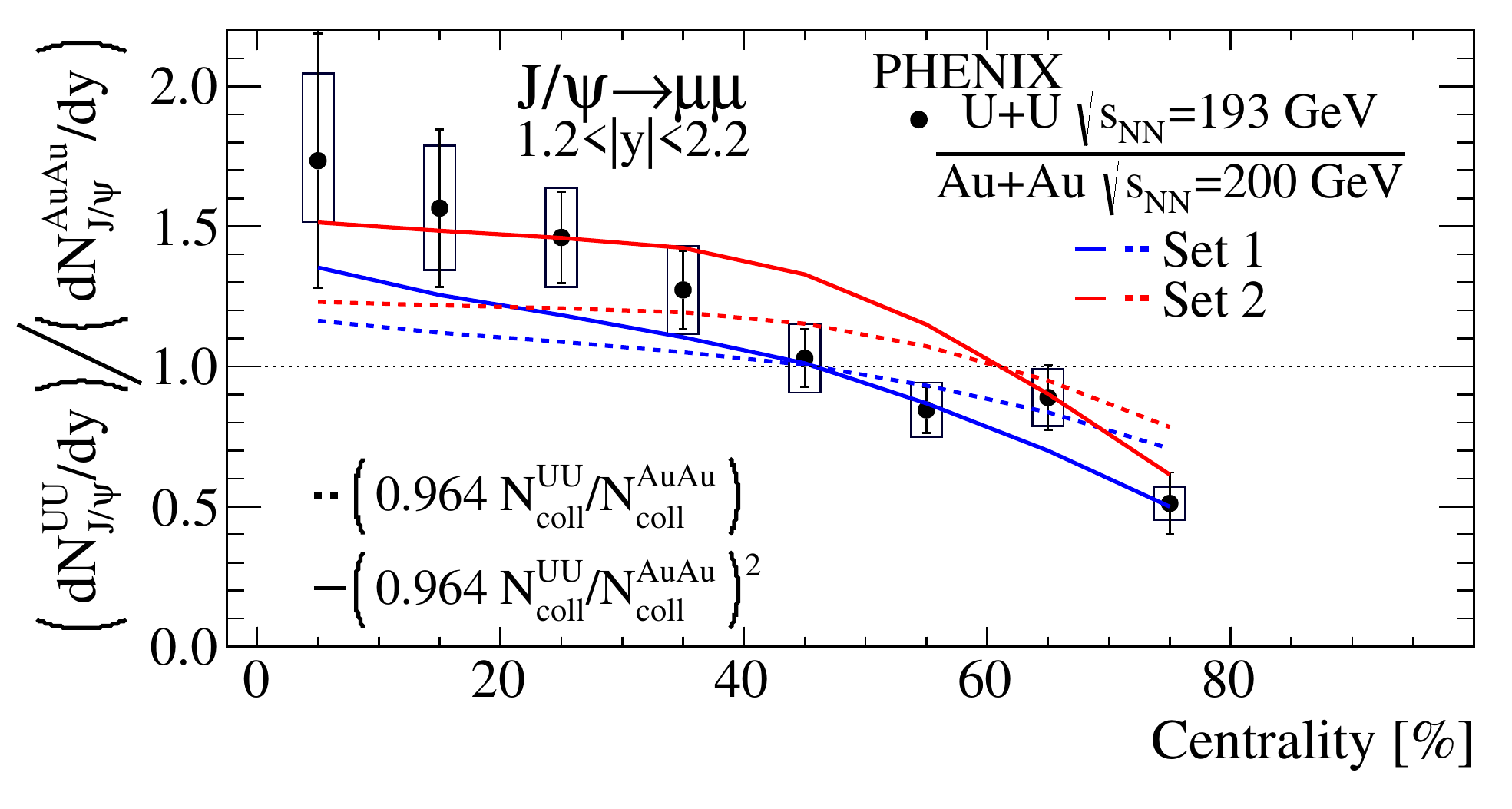}	
    \centering
    \caption{\label{fig:BdNdyratio}
      (Color online) The ratio of the invariant yield for U$+$U to that for Au$+$Au, as a  function of collision centrality. 
      The curves show how the ratio would vary with centrality if the invariant yield scaled with  \Ncoll (dashed curve) or with $\Ncoll^2$ (solid curve). 
      The \Ncoll ratio curves obtained from the U$+$U Glauber model calculation using Woods-Saxon parameter set 1~\cite{Masui:2009qk} 
      are shown in blue, those using parameter set 2~\cite{Shou:2014eya} are shown in red.
      The \Ncoll value for \uu is multiplied by a factor of 0.964 to account for the decrease
      of the charm cross section for \pp collisions from 200~GeV to 193 
GeV collision energy. 
  } 
\end{figure}

The values of $R^{UU}_{\rm AuAu}$ are plotted in 
Figure~\ref{fig:Ratio_RAAN2_XSect}, and summarized in 
Table~\ref{tbl:RUUAuAu}. For the \Ncoll values from set 2, the relative 
nuclear modification falls below one for collisions in the 40\%--60\% region, 
but rises to one for the most central collisions. For \Ncoll values from 
set 1, the relative nuclear modification is slightly above 1 for central collisions and 
approximately one across the remainder of the 
centrality range. These results suggest that the invariant yield scales 
with $\Ncoll^2$ for the most central collisions, at least.

\begin{table*}
  \caption{\label{tbl:BdNdy} 
The invariant yield for \uu collisions (averaged over forward 
(1.2\mylessthan$y$\mylessthan2.2) and backward 
(-2.2\mylessthan$y$\mylessthan-1.2) rapidity), the invariant yield for 
\auau collisions, and the ratio of invariant yields plotted in 
Fig.~\ref{fig:BdNdyratio}, all as a function of centrality. The first and 
second uncertainties listed represent Type-A (statistical uncertainties 
plus point-to-point systematic uncertainties from the yield extraction) 
and Type-B uncertainties, respectively (see text for definitions). There 
is no Type-C (global) systematic uncertainty for the invariant yield.
}
  \begin{ruledtabular} \begin{tabular}{ccccccccc}
&  Centrality &   $10^6 B_{\mu\mu}dN^{UU}/dy$                            &  $10^{6} B_{\mu\mu}dN^{\rm AuAu}/dy$                            &  $\frac{dN^{UU}/dy}{dN^{\rm AuAu}/dy}$ \\
\hline
&  0\%--10\%     & 182.7$\pm$   35.1 $\pm$    15.7  &  105.3 $\pm$ 18.8 $\pm^{16.6}_{9.72}$  &  1.73 $\pm$ 0.46 $\pm^{0.31}_{0.22}$   \\  
& 10\%--20\%   &  144.5 $\pm$    19.8 $\pm$  12.4  &  92.3 $\pm$ 10.7 $\pm^{10.6}_{10.3}$    &  1.57 $\pm$ 0.28 $\pm$ 0.22        \\
& 20\%--30\%   &  110.6 $\pm$     8.2 $\pm$   9.5    &  75.8 $\pm$ 6.3 $\pm$ 6.4                       &   1.46 $\pm$ 0.16 $\pm$ 0.18      \\
& 30\%--40\%   &  72.4 $\pm$     6.3 $\pm$     6.2    & 56.8 $\pm$ 3.8 $\pm$ 5.1                        &   1.27 $\pm$ 0.14 $\pm$ 0.16       \\
& 40\%--50\%   &  41.7 $\pm$     3.4 $\pm$     3.6    &  40.5 $\pm$ 2.4 $\pm$ 3.4                       &   1.03 $\pm$ 0.10 $\pm$ 0.12       \\
& 50\%--60\%   &  23.8 $\pm$     2.0 $\pm$     2.0    &  28.2 $\pm$ 1.4 $\pm$ 2.1                       &   0.84 $\pm$ 0.08 $\pm$ 0.10       \\
& 60\%--70\%   & 11.9 $\pm$     1.4 $\pm$     1.0     &  13.4 $\pm$ 0.8 $\pm$ 1.0                       &   0.89 $\pm$ 0.12 $\pm$ 0.10       \\
& 70\%--80\%   & 4.0 $\pm$     0.8 $\pm$       0.4     &  7.9 $\pm$ 0.54 $\pm$ 0.60                     &   0.51 $\pm$ 0.11 $\pm$ 0.06       \\
  \end{tabular} \end{ruledtabular}
\end{table*}

As a different way of looking at the data, we show in 
Figure~\ref{fig:BdNdyratio} the ratio of the invariant yields for \uu and 
\auau, taken from Table~\ref{tbl:BdNdy}. These do not depend on \Ncoll. 
For comparison, in Figure~\ref{fig:BdNdyratio} we also present curves 
showing how the ratio would depend on centrality if it scaled with \Ncoll 
(dashed curve) or with $\Ncoll^2$ (solid curve). As in 
Equation~\ref{eqn:RUUAuAu}, the values of \Ncoll for \uu are multiplied by 
0.964 to account for the difference in cross section for \jpsi production 
in \pp collisions at 193~GeV and 200~GeV collision energy. When \Ncoll 
values from set 2 are used, the measured ratios for midcentral collisions 
lie at or below the ratio of \Ncoll values, but as the collision 
centrality increases the data points move above the ratio of \Ncoll values 
until, for the most central collisions, they favor the $\Ncoll^2$ curve. 
When \Ncoll values from set 1 are used, the data slightly favor the trend 
of the $\Ncoll^2$ curve across the centrality range.

\begin{table}
  \caption{\label{tbl:RUUAuAu} 
The relative nuclear-modification factor for \uu and \auau collisions (see 
Equation~\ref{eqn:RUUAuAu}) (averaged over forward 
(1.2\mylessthan$y$\mylessthan2.2) and backward 
(-2.2\mylessthan$y$\mylessthan-1.2) rapidity). The values derived from the 
deformed Woods-Saxon parameter sets 1~\cite{Masui:2009qk} and 
2~\cite{Shou:2014eya} are shown. The first and second uncertainties listed 
represent Type-A (statistical uncertainties plus point-to-point systematic 
uncertainties from the yield extraction) and Type-B uncertainties, 
respectively (see text for definitions). There is no Type-C (global) 
systematic uncertainty.
}
  \begin{ruledtabular} \begin{tabular}{ccccccccc}
&  Centrality &    $R^{UU}_{\rm AuAu}$ (set 1)   & $R^{UU}_{\rm AuAu}$ (set 2)  \\
\hline
&  0\%--10\% & 1.29 $\pm$ 0.34 $\pm^{0.23}_{0.16} $ &  1.13 $\pm$ 0.30 $\pm^{0.20}_{0.14}$ \\  
& 10\%--20\%   & 1.27 $\pm$ 0.23 $\pm$ 0.18     &   1.04 $\pm$ 0.19 $\pm$ 0.15 \\
& 20\%--30\%   &  1.26 $\pm$ 0.14 $\pm$ 0.15    &   0.96 $\pm$ 0.11 $\pm$ 0.12 \\
& 30\%--40\%   &  1.21 $\pm$ 0.13 $\pm$ 0.15    &   0.86 $\pm$ 0.09 $\pm$ 0.11 \\
& 40\%--50\%   &  1.12 $\pm$ 0.11 $\pm$ 0.13    &   0.72 $\pm$ 0.07 $\pm$ 0.09 \\
& 50\%--60\%   &  1.03 $\pm$ 0.10 $\pm$ 0.12    &   0.64 $\pm$ 0.06 $\pm$ 0.07 \\
& 60\%--70\%   &  1.25 $\pm$ 0.16 $\pm$ 0.14    &   0.83 $\pm$ 0.11 $\pm$ 0.10 \\
& 70\%--80\%   &  0.96 $\pm$ 0.21 $\pm$ 0.11    &   0.64 $\pm$ 0.14 $\pm$ 0.07 \\
  \end{tabular} \end{ruledtabular}
\end{table}

These comparisons of the ratios of the \uu and \auau invariant yields with 
\Ncoll values derived from both deformed Woods-Saxon parameter sets 
suggests that, for central collisions, increased $c\bar{c}$ coalescence is more 
important than stronger suppression due to increased energy density. When 
the \Ncoll values from set 2 are used, the peripheral and midcentral data 
suggest that the increased suppression due to increased energy density and 
increased coalescence due to larger underlying charm yields approximately 
cancel each other. When the \Ncoll values from set 1 are used, the 
peripheral and midcentral ratios are consistent with either \Ncoll or 
$\Ncoll^2$ scaling.

			\section{Summary}

We have presented measurements of the \pt integrated invariant yield $dN/dy$ and 
nuclear modification \raa for \jpsi production in \uu collisions at 
$\sqrt{s_{NN}}$ = 193~GeV, and compared them with existing data for \auau 
collisions at $\sqrt{s_{NN}}$ = 200~GeV. In addition to comparing the 
invariant yields and nuclear modification, \raa, for the two systems, we 
have combined them to form the relative nuclear modification, 
$R^{UU}_{\rm AuAu}$~\cite{Kikola:2011zz}. The relative nuclear modification 
(Equation~\ref{eqn:RUUAuAu}) was proposed as a way to eliminate the 
systematic uncertainties associated with the formation of \raa, and to 
cancel most of the systematic uncertainties associated with the number of 
binary nucleon collisions, \Ncoll. It is designed to have a value of one 
if the \jpsi cross section is dominated by $c\bar{c}$ coalescence. We have 
also compared the ratio of the invariant yields for \uu and \auau with the 
ratio of \Ncoll values and the ratio of $\Ncoll^2$ values for \uu and 
\auau.

We discussed the effect of using two different parameterizations for the 
deformed Woods-Saxon distribution on the estimates of \Ncoll for \uu 
collisions. A recently proposed method for estimating the Glauber model 
parameters (set 2)~\cite{Shou:2014eya} leads to a smaller surface 
diffuseness for U, and thus larger values of \Ncoll, than a conventional 
estimate (set 1)~\cite{Masui:2009qk}. We presented \raa values and 
relative nuclear modification values obtained using \Ncoll values from 
both deformed Woods-Saxon parameter sets.

For both sets of \Ncoll values the \raa for \uu is found to be less 
suppressed than for \auau in central collisions that have a similar number 
of participants. The relative nuclear modification is found to be one for 
the most central collisions for both \Ncoll sets, but for set 2 it falls 
below one for mid-peripheral and peripheral collisions. When the ratios of 
invariant yields are compared with the ratios of \Ncoll and $\Ncoll^2$ 
values, they are found to show a slight preference for $\Ncoll^2$ scaling 
for central collisions.

For both sets of \Ncoll values the behavior is consistent with a picture 
in which, for central collisions, the increase in \jpsi yield for \uu due 
to $c\bar{c}$ coalescence becomes more important than the decrease in 
yield due to increased energy density. For \Ncoll values from set 1, the 
results are consistent with both \Ncoll and $\Ncoll^2$ scaling for 
mid-peripheral collisions. For \Ncoll values from set 2, the results 
suggest that in the 40\%--60\% centrality range the increased suppression due 
to higher energy density in \uu collisions is more important than the 
increased \jpsi yield due to coalescence caused by the higher underlying 
charm production.


	\section*{ACKNOWLEDGMENTS}   

We thank the staff of the Collider-Accelerator and Physics
Departments at Brookhaven National Laboratory and the staff of
the other PHENIX participating institutions for their vital
contributions.  We acknowledge support from the
Office of Nuclear Physics in the
Office of Science of the Department of Energy, the
National Science Foundation, Abilene Christian University
Research Council, Research Foundation of SUNY, and Dean of the
College of Arts and Sciences, Vanderbilt University (U.S.A),
Ministry of Education, Culture, Sports, Science, and Technology
and the Japan Society for the Promotion of Science (Japan),
Conselho Nacional de Desenvolvimento Cient\'{\i}fico e
Tecnol{\'o}gico and Funda\c c{\~a}o de Amparo {\`a} Pesquisa do
Estado de S{\~a}o Paulo (Brazil),
Natural Science Foundation of China (People's Republic of~China),
Croatian Science Foundation and
Ministry of Science, Education, and Sports (Croatia),
Ministry of Education, Youth and Sports (Czech Republic),
Centre National de la Recherche Scientifique, Commissariat
{\`a} l'{\'E}nergie Atomique, and Institut National de Physique
Nucl{\'e}aire et de Physique des Particules (France),
Bundesministerium f\"ur Bildung und Forschung, Deutscher
Akademischer Austauschdienst, and Alexander von Humboldt Stiftung 
(Germany),
National Science Fund, OTKA, K\'aroly R\'obert University College 
(Hungary)
Department of Atomic Energy and Department of Science and Technology 
(India),
Israel Science Foundation (Israel),
Basic Science Research Program through NRF of the Ministry of Education 
(Korea),
Physics Department, Lahore University of Management Sciences (Pakistan),
Ministry of Education and Science, Russian Academy of Sciences,
Federal Agency of Atomic Energy (Russia),
VR and Wallenberg Foundation (Sweden),
the U.S. Civilian Research and Development Foundation for the
Independent States of the Former Soviet Union,
the Hungarian American Enterprise Scholarship Fund,
and the US-Israel Binational Science Foundation.

  
 
%
 
\end{document}